\RequirePackage{ifpdf}
\documentclass[a4paper,11pt]{article}
\pdfoutput=1 

\usepackage{jheppub}
\usepackage{amsmath,latexsym}
\usepackage{longtable}
\usepackage{supertabular}
\usepackage{color}
\usepackage[small]{subfigure}

\usepackage[T1]{fontenc}

\newcommand{\df}{\mathrm{d}}
\newcommand{\as}{\alpha_{\rm S}}

\title{\boldmath Oriented Event Shapes at N${}^3$LL + $\mathcal{O}(\as^2)$}

\preprint{ \begin{flushright} IFIC/13-40\\
LPN13-044
\end{flushright}\vspace*{-1cm}}

\author[1]{Vicent Mateu\note{Corresponding author.}}
\author{and  Germ\'an Rodrigo}
\affiliation{Instituto de F\'\i sica Corpuscular, \\ Universitat de Val\`encia --
Consejo Superior de Investigaciones Cient\'\i ficas,  \\
Parc Cient\'{\i}fic, E-46980 Paterna (Valencia), Spain.}

\emailAdd{mateu@ific.uv.es}
\emailAdd{german.rodrigo@csic.es}
\abstract{We analyze oriented event-shapes in the context of
Soft-Collinear Effective Theory (SCET) and in fixed-order
perturbation theory. Oriented event-shapes are distributions of event-shape variables which
are differential on the angle $\theta_T$ that the thrust axis forms with the electron--positron
beam. We show that at any order in perturbation theory and for any event shape,
only two angular structures
can appear: $F_0 = 3/8\,(1 + \cos^2\theta_T)$ and $F_1 = (1 - 3\,\cos^2\theta_T)$.
When integrating over $\theta_T$ to recover the more familiar event-shape distributions,
only $F_0$ survives. The validity of our proof goes beyond perturbation theory, and hence
only these two structures are present at the hadron level. The proof also carries over
massive particles. Using SCET techniques we show that singular terms can only arise in the
$F_0$ term. Since only the hard function is sensitive to the orientation of the thrust axis,
this statement applies also for recoil-sensitive variables such as Jet Broadening.
We show how to carry out resummation of the singular terms at N${}^3$LL 
for Thrust, Heavy-Jet Mass, the sum of the Hemisphere Masses and \mbox{$C$-parameter} 
by using existing computations in SCET. We also compute the fixed-order
distributions for these event-shapes at $\mathcal{O}(\as)$ analytically and at
$\mathcal{O}(\as^2)$ with the program Event2.} 

\keywords{QCD, perturbative corrections}

\begin{document}
\maketitle
\flushbottom

\section{Introduction}
\label{sec:introduction}

The experimental collaborations at LEP collected very precise data for event-shape
distributions at the $Z$-pole energy (LEP1) and also at higher energies (LEP2) with
somewhat larger uncertainties. These data, along with measurements at lower
energies from other experiments, have been recently analyzed for the thrust
distribution~\cite{Abbate:2010xh,Gehrmann:2012sc} and 
moments~\cite{Gehrmann:2009eh,Abbate:2012jh} using higher-order resummation, three-loop
matrix elements and analytic methods to parametrize non-perturbative power 
corrections. These analyses found rather low values of the strong coupling
$\as$ with quite small uncertainties, which are in disagreement with the
world average~\cite{PDG:2012}~\footnote{See \cite{Bethke:2011tr} for an overview
of recent $\as$ determinations.}. Further analysis for $C$-parameter and
Heavy-Jet Mass are on the way~\cite{Hoang:2013,Hoang:2013a}.

These analyses, and the vast majority of the experimental activity, have focused
only on angular-averaged event-shapes. By averaged we want to stress the fact
that there was no information recorded on the orientation of the event with respect
to the beam axis. The DELPHI collaboration delivered~\cite{Abreu:2000ck} very
accurate measurements of eighteen oriented infrared and collinear safe observables
at the $Z$-peak from $1.5$ million collected events. Oriented
distributions are differential in the event-shape variable, as usual, but also on
the polar angle formed by the thrust axis and the beam direction, see 
Fig.~\ref{fig:thrust-axis} (events are symmetric
with respect to the azimuthal angle and hence its dependence is integrated).
They also presented a very accurate determination of $\as$ 
by performing two-parameter fits to $\as$ and $x_\mu$, the square of
the ratio between the renormalization scale $\mu$ and the center-of-mass energy $Q$.
Data was compared to ${\mathcal O}(\as^2)$ 
theoretical computations~\cite{Catani:1996jh,Catani:1996vz,Rodrigo:1997gy}, 
neglecting QED corrections.
Even though they analyzed the effect of NLL
resummation, their main analysis relies on fixed-order perturbation theory.

The OPAL collaboration also performed oriented measurements 
with respect to the thrust axis for both the total cross section and the thrust distribution 
from $2.1$ million collected events~\cite{Abbiendi:1998at}. Data were presented at the
hadron and parton level, corrected with Monte Carlo generators.

In this article we take a first step towards an analysis which uses the most recent
theoretical developments: ${\mathcal O}(\as^3)$ matrix elements 
\cite{GehrmannDeRidder:2007bj,
GehrmannDeRidder:2007hr,Weinzierl:2008iv,Weinzierl:2009ms} and higher-order log
resummation. We also show how to treat oriented distributions in a more systematic
fashion, demonstrating that there is only one additional piece of information
as compared to averaged event-shapes. We compute this new piece analytically at
${\mathcal O}(\as)$ and extract it numerically at ${\mathcal O}(\as^2)$
using Event2~\cite{Catani:1996jh,Catani:1996vz}. 
We also proof that singular logs can only occur in the angular averaged piece.
Fixed-order computations of oriented
event-shapes at $\mathcal{O}(\alpha_s)$ fully differential and for thrust
were presented in Ref.~\cite{Lampe:1992au}, together with 
a numerical determination of the $\mathcal{O}(\alpha_s^2)$ contribution
to the integrated longitudinal cross section;
an SCET factorization theorem for the longitudinal piece of the thrust
distribution was derived in \cite{Hagiwara:2010cd}, achieving NLL resummation
(this term is subleading in the SCET counting). 
In this article we derive the singular behavior of the various angular structures
directly from SCET, without assuming a splitting in transverse and longitudinal parts.

A thorough comparison to experimental data including ${\mathcal O}(\as^3)$
matrix elements, QED and bottom-quark mass effects along with non-perturbative
power corrections is left for future work.

The event-shape variables that we consider in this article are:
\begin{enumerate}
 \item Thrust~\cite{Farhi:1977sg}\,:\\
 Thrust is defined as the sum of the absolute values of the projections of each particle three-momentum 
 on a given direction $\hat n$. This direction is chosen such that the sum is maximized, and the result is then
 normalized to the sum of the magnitudes of the three-momenta of all particles~\footnote{If one normalizes
to the center-of-mass energy instead, then the resulting event-shape is called
\mbox{2-jettiness}~$\tau_2$~\cite{Stewart:2010tn}. For massless particles $\tau = \tau_2$,
 see \cite{Mateu:2012nk}.}
 \begin{align}\label{eq:thrustDef}
T = 1- \tau = \dfrac{1}{\sum_i |\vec{p}_i|}\max_{\hat n}\sum_i|{\hat n}\cdot {\vec p}_i|\,.
 \end{align}
It is customary to define $\tau = 1 - T$, since as $\tau\to 0$ one reaches the dijet limit. For
simplicity we will use the name thrust also for $\tau$.
 \item $C$-parameter~\cite{Parisi:1978eg,Donoghue:1979vi}\,:\\
 Unlike thrust, $C$-parameter does not require any minimization procedure.
 It is defined in terms of the eigenvalues of the linearized momentum tensor.
 It can be written as a double sum over the three-momenta magnitudes and relative angle
 of pairs of particles, normalized analogously to thrust~\footnote{For a detailed derivation
 of how to express the eigenvalue in terms of a double sum the reader is referred
 to \cite{Hoang:2013a}.}
 \begin{align} \label{eq:CparamDef}
 C=\frac{3}{2} \frac{\sum_{i,j} | \vec{p}_i | | \vec{p}_j | \sin^2 \theta_{ij}}
 {\left( \sum_i | \vec{p}_i | \right)^2}\,.
 \end{align}
 \item Hemisphere Masses~\cite{Clavelli:1979md, Chandramohan:1980ry, Clavelli:1981yh}\,:\\
 The plane normal to the thrust axis $\hat n$ defines two separated hemispheres  
 (see Fig.~\ref{fig:thrust-axis}) which shall be denoted by $a$ and $b$. The
 hemisphere masses are defined as the square of the total four-momentum in each hemisphere
 \begin{align} \label{eq:HemiDef}
 S_{a,b} = \bigg(\sum_{i\in a,b}p_i^\mu\bigg)^2\,,
 \end{align}
 One can define two dijet event-shapes out of the hemisphere masses, Heavy-Jet Mass 
 $\rho$~\footnote{Heavy-Jet Mass is also represented by $\rho_H$.}
 and the sum of the hemisphere masses $\rho_S$, and the non-dijet event-shapes Light-Jet Mass $\rho_L$
 and the absolute value of the difference of the hemisphere masses $\rho_D$, all of  
 them normalized to the square of the center-of-mass energy $Q^2$
 \begin{alignat}{2} \label{eq:rhoDef}
 &\rho = \dfrac{1}{Q^2}\max(S_a,S_b)\,, &&\quad\quad\rho_L = \dfrac{1}{Q^2}\min(S_a,S_b)\,,\\
 &\rho_S = \dfrac{S_a + S_b}{Q^2} = \rho + \rho_L\,, &&
 \quad\quad\rho_D = \dfrac{|S_a - S_b|}{Q^2} = \rho - \rho_L\,.\nonumber
 \end{alignat}
 The dijet configuration is achieved when both hemisphere masses are small. Heavy-Jet Mass is a dijet
 event-shape because when $\rho$ is small both hemisphere masses have to be small. The same goes true for
 $\rho_S$. However $\rho_L$ can be small when one hemisphere mass is big while the other is small,
 and $\rho_D$ can be small for big hemisphere masses but of similar size.
 There is a relation between the hemisphere masses and thrust and 2-jettiness:
\begin{align}\label{eq:MassesThrust}
\tau & = 1 - \dfrac{Q}{\sum_i |\vec{p}_i|}\sqrt{1 - 2\,\rho_S + \rho_D^2}~,\\
\tau_2 & = 1 - \sqrt{1 - 2\,\rho_S + \rho_D^2}~.\nonumber
\end{align}
Expanding at leading order in the dijet limit one has $\tau_2 = \rho_S + {\mathcal O}(S_{a,b}^2)$.
The mass of a hemisphere is zero when it is populated by a single massless particle only, hence
$\rho_L$ is zero for events with two or three massless particles. Plugging $\rho_L = 0$ into Eqs.~(\ref{eq:rhoDef})
and (\ref{eq:MassesThrust}) one finds that
thrust, $\rho$, $\rho_S$ and $\rho_D$ are identical for two and three (massless) particles.
Hence the ${\mathcal O}(\as)$ fixed-order computation of Sec.~\ref{sec:LO} is the same for the
three event-shapes, and we
will present the results for thrust and $C$-parameter only. At NLO we will present results for $\tau$, $\rho$ and $\rho_S$, as well as for
$C$-parameter.
\end{enumerate}

This paper is organized as follows: In Sec.~\ref{sec:SCET} we derive an
all-orders factorization theorem for oriented event-shapes in Soft-Collinear
Effective Theory (SCET) \cite{Bauer:2000ew, Bauer:2000yr, Bauer:2001ct, Bauer:2001yt,
Bauer:2002nz}, which permits resummation at N${}^3$LL; in
Sec.~\ref{sec:LO} we compute the ${\mathcal O(\as)}$ fixed-order
distribution for oriented event-shapes;
in Sec.~\ref{sec:Angular} we show that to all orders in
perturbation theory and even at the hadronic level, only two angular structures
can arise; in Sec.~\ref{sec:NLO} we determine the ${\mathcal O(\as^2)}$
fixed-order distribution for some oriented distributions using Event2, along
with the oriented total cross-section; conclusions and outlook can be found in
Sec.~\ref{sec:conclusions}.

\section{SCET Factorization Theorem for Oriented Event-Shapes}
\label{sec:SCET}
In this section we derive the SCET factorization theorem for the most singular terms in the oriented 
event-shape distributions. Factorized expressions in SCET are tremendously useful as they a) allow to carry out
resummation of large logarithms to very high accuracy through Renormalization Group Evolution (RGE);
as is the case of thrust \cite{Becher:2008cf,Abbate:2010xh}, Heavy-Jet Mass \cite{Chien:2010kc,Hoang:2013},
$C$-parameter \cite{Hoang:2013a} at N${}^3$LL, Jet Broadening~\cite{Rakow:1981qn}
at N${}^2$LL \cite{Becher:2011pf,Becher:2012qc},
or angularities \cite{Berger:2003iw,Hornig:2009vb} at NLL\,\footnote{The alternative to factorization
theorems are the classic exponentiation techniques of Ref.~\cite{Catani:1992ua},
which works out explicitly the cases of thrust and Heavy Jet Mass. These results
were extended to $C$-parameter in Ref.~\cite{Catani:1998sf}. For Jet Broadening the LL
resummation was carried out in Ref.~\cite{Catani:1992jc} whereas the NLL extension was completed in
Ref.~\cite{Dokshitzer:1998kz}. The NLL resummation of any event shape has been automatized in 
Refs.~\cite{Banfi:2003je,Banfi:2004yd}. Resummation in this formalism has so far only been carried
out to NLL order, with the exception of \cite{deFlorian:2004mp} which achieves N${}^2$LL};
b) simplify the calculation of the necessary ingredients (matrix elements and anomalous dimensions);
c) confine non-perturbative effects to specific
functions~\cite{Bauer:2002ie,Bauer:2003di,Mateu:2012nk}.
Factorization of event-shape cross-sections
was first performed in QCD in \cite{Korchemsky:1998ev,Korchemsky:1999kt,Korchemsky:2000kp}
and was followed by Effective Field Theory techniques in \cite{Fleming:2007qr,Schwartz:2007ib,Hoang:2007vb}.
Recently subleading corrections to the thrust distribution have been derived in SCET \cite{Freedman:2013vya}.

Our proof closely follows the derivation given in Ref.~\cite{Bauer:2008dt} and actually requires only minimal
modifications. Let $p_1$ and $p_2$ be the four-momenta of the incoming electron and positron, respectively,
and let $Q$ be the center-of-mass energy.~\footnote{Throughout this article we use $p_i$ to denote
initial state momenta (that is incoming electron and positron momenta), and $q_i$ to denote final
state momenta (that is quark, gluon or hadron momenta).}
The leptonic tensor involved in the process $e^+e^-\to\rm hadrons$ can be split into vector and
axial contributions, and is given by~\footnote{Unless otherwise stated, throughout this article we
ignore P-violating terms proportional
to $\cos\theta_T$ since the thrust axis does not allow to distinguish $\theta_T$
from $\pi - \theta_T$ with light jets.} 
\begin{align}\label{eq:lepton-tensor}
L_{V \, (A)}^{\mu \nu} & = \dfrac{16\pi^2\alpha^2}{Q^4}\,
\mathcal{V}^{\mu\nu}\,L_{V \, (A)}~,\\
\mathcal{V}^{\mu\nu} & = p_1^\mu p_2^\nu + p_2^\mu p_1^\nu-\dfrac{Q^2}{2}\,g^{\mu\nu}~,\nonumber
\end{align}
where $L_{V \, (A)}$ are electroweak factors which can be found for instance in Eq.~(9) of Ref.~\cite{Bauer:2008dt}.
Considering only QED interactions $L_V = \sum_f Q_f^2$ and $L_A = 0$, being $f$ the quark flavor.
In full QCD an event shape can be written as
\begin{align}~\label{eq:QCD}
\dfrac{\df \sigma}{\df e} & = \dfrac{1}{2Q^2}\!\int \df^4x\, e^{i q\cdot x}\sum_{i=V,A}
L^i_{\mu\nu}\,\langle \, 0\,|\, j_i^{\mu\,\dagger}(x)\, \delta(e - {\hat e})\,
j_i^\nu(0)\,|\,0\, \rangle\,,\\
j_i^\mu & = \bar{q}_f\,\Gamma_i^\mu\,q_f\,,\quad
\Gamma_V^\mu = \gamma^\mu\,,\quad
\Gamma_A^\mu = \gamma^\mu\gamma_5\,,\nonumber
\end{align}
where $\hat e$ is an operator that when acting on a state pulls out as an eigenvalue
the value of the event shape $e$\,: $\hat e\,|\,X\, \rangle = e(X)\,|\,X\, \rangle$,
and $i = V, A$ labels vector and axial currents. In the second line of
Eq.~(\ref{eq:QCD}) there is an implicit sum over the color of the quarks. In the rest
of the proof we omit sums over the quark flavors, which can be trivially inferred.
The next step is to match the QCD current $j_i^\mu$ onto SCET:
\begin{align}~\label{eq:matching}
j_i^\mu & = \sum_{\hat n}\sum_{\tilde{p}_1,\tilde{p}_2}
C_{n,\bar{n}}(\tilde{p}_1,\tilde{p}_2,\mu)\,
\mathcal{O}_{n\bar{n}}(x;\tilde{p}_1,\tilde{p}_2)\,,\\
\mathcal{O}_{n\bar{n}}(x;\tilde{p}_1,\tilde{p}_2) & =
e^{i(\tilde{p}_1-\tilde{p}_2)\cdot x}\,
\bar{\chi}_{n,\tilde{p}_1}(x)Y_n(x)\Gamma_i^\mu
\overline{Y}_{\bar n}(x)\chi_{{\bar n},{\tilde p}_2}\,.
\nonumber
\end{align}
$n^\mu$ and ${\bar n}^\mu$ are light-like vectors in the direction of the
primary quark and anti-quark four-momenta. More specifically, if $\hat n$
is the direction of the thrust axis, then
$n^\mu = (1,{\hat n})$, ${\bar n}^\mu = (1,-\,{\hat n})$, such that $n^2 = \bar{n}^2 = 0$
and $n\cdot \bar{n} = 2$. Since the SCET spinors have only two components one can
simplify the Dirac matrices to $\Gamma_V^\mu = \gamma_\perp^\mu$ and
$\Gamma_A^\mu = \gamma_\perp^\mu\gamma_5$, with $\gamma_\perp$ the projection of the
Dirac matrices into the $x$\,--\,$y$ plane: $\gamma_\perp^\mu \,=\, \gamma^\mu \,-\,
n\!\!\!\slash\, \bar{n}^\mu/2 \,-\, \bar{n}\!\!\!\slash\, n^\mu/2$. $Y_n$ and
$\overline{Y}_{\bar{n}}$ are path- and anti-path orderer soft Wilson lines,
respectively, in the corresponding light-like directions. $\chi$ represents a
jet field, which is the product of a collinear quark field and a collinear Wilson
line, making the combination collinear gauge invariant. $\tilde{p}_j$ are
large label momenta (we use here the label formalism).
In Eq.~(\ref{eq:matching}) we have used a field redefinition \cite{Bauer:2001yt}
that decouples soft and collinear degrees of freedom at leading order in the SCET
Lagrangian. This is the first step to factorize the cross section.

Next one replaces Eq.~(\ref{eq:matching}) into (\ref{eq:QCD}) and requires
label momentum conservation. This fixes
$\bar{n}\cdot \tilde{p}_1 = - \, n\cdot \tilde{p}_2 = Q$ and makes the label 
transverse momenta zero. Then the hard function is defined as
$H(Q,\mu) \equiv |C_{n,\bar{n}}(Q,Q,\mu)|^2$.
In the next step towards factorization the event shape delta 
function is decomposed as follows:
\begin{align}
\delta(e - {\hat e}) =\! \int \df e_n\,\df e_{\bar n}\,\df e_s\,
\delta(e_n - \hat{e}_n)\, \delta(e_{\bar n} - \hat{e}_{\bar n})\,
\delta(e_s - \hat{e}_s)\,\delta(e - e_n - e_{\bar n} - e_s)\,,
\end{align}
where $\hat{e}_n$, $\hat{e}_{\bar n}$ and $\hat{e}_s$ are event-shape operators
acting only on the $n$-, $\bar{n}$-collinear and soft sectors, respectively.
Then, color conservation and the Fierz identity for Dirac
matrices are used to place next to one another fields belonging to the various sectors. After some
manipulations of the various matrix elements, as detailed in Ref.~\cite{Bauer:2008dt},
one arrives at
\begin{align}
\dfrac{\df \sigma}{\df e} & = K_0\, H(Q,\mu)\sum_{\hat{n}}
\dfrac{\df^2\vec{k}_\perp}{2(2\pi)^2}\!\int\!\df e_n\,\df e_{\bar n}\,\df e_s\,
\delta(e - e_n - e_{\bar n} - e_s)\,J_n(e_n,\mu)\,J_{\bar n}(e_{\bar n},\mu)
\,S_e(e_s,\mu)\,,\nonumber\\
K_0 & = \dfrac{N_C}{2Q^2}\!\sum_{i=V,A} L_i^{\mu\nu}\,
{\rm Tr}\Big[\dfrac{\bar{n}\!\!\!\slash}{2}\,\Gamma_i^\mu\,
\dfrac{n\!\!\!\slash}{2}\,\overline{\Gamma}_i^\nu\Big]\,,
\label{K0trace}
\end{align}
where $J$ and $S$ are the jet and soft functions, respectively. Following Ref.~\cite{Fleming:2007qr}
one can work out together the sum over $\hat n$ and the integral measure over the transverse momentum
to find
\begin{align}
\sum_{\hat{n}}\dfrac{\df^2\vec{k}_\perp}{2(2\pi)^2} =
\dfrac{Q^2}{32\pi^2}\,\df \Omega\,.
\end{align}
\begin{figure}[t!]
\begin{center}
\includegraphics[width=6cm]{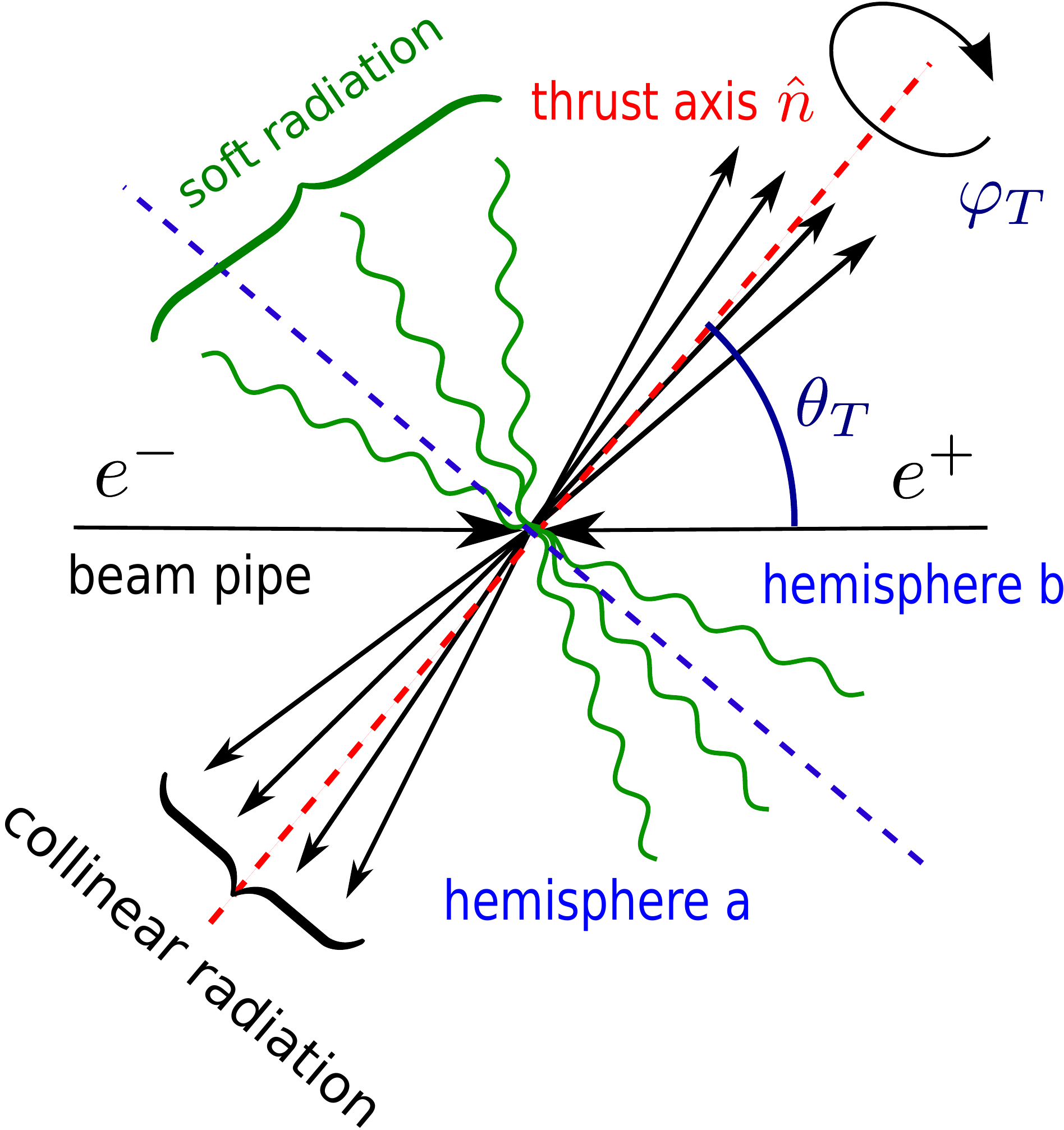}
\caption{Schematic dijet event. The thrust axis is depicted as a dashed red line.
Collinear particles are represented as black
arrows around the thrust axis. Soft radiation appears as green wiggly lines
and can be emitted in central regions of the phase-space. The two hemispheres $a$ and $b$ are separated by
a plane normal to the thrust axis, and in this figure it appears as a blue dashed
line. $\theta_T$ is defined as the angle between the beam (for
example the initial-state electron) and the thrust axis. $\varphi_T$ is the orientation of the
event in the azimuthal direction around the thrust axis. We average over that angle.
\label{fig:thrust-axis}}
\end{center}
\end{figure}
Here $\Omega$ refers to the orientation of the thrust axis with respect to the beam, as illustrated in
Fig.~\ref{fig:thrust-axis}.  Finally, calculating the trace in 
Eq.~(\ref{K0trace}) and collecting some factors, we find that the most singular contribution 
to the differential angular distribution of the event-shape $e$ is
\begin{align}
\dfrac{1}{\sigma_0}\dfrac{\df \sigma_{\rm S}}{\df \Omega\,\df e} = \dfrac{3}{16\pi}\dfrac{1}{Q^2}
(n_\mu {\bar n}_\nu + n_\nu {\bar n}_\mu - 2g_{\mu\nu})\mathcal{V}^{\mu\nu}
\dfrac{1}{\sigma_0}\dfrac{\df \sigma_{\rm S}}{\df e}\,,
\end{align}
where
\begin{align}
\sigma_0 = \dfrac{4\pi\alpha^2N_C}{3Q^2}\,(L_V + L_A)\,,
\end{align}
is the Born cross-section. It is straightforward to compute
$(n_\mu {\bar n}_\nu + n_\nu {\bar n}_\mu - 2g_{\mu\nu})\mathcal{V}^{\mu\nu} = Q^2 (1 + \cos^2\theta_T)$,
which leads to our final result:
\begin{align}\label{eq:final-SCET}
\dfrac{1}{\sigma_0}\dfrac{\df \sigma_{\rm S}}{\df\!\cos\theta_T\,\df e} = 
\dfrac{3}{8}\,(1 + \cos^2\theta_T)\,\dfrac{1}{\sigma_0}\dfrac{\df \sigma_{\rm S}}{\df e}\,.
\end{align}
The form of Eq.~(\ref{eq:final-SCET}) is actually very simple: the most singular terms of the oriented
event-shape distribution inherit the angular dependence of the lowest order process $e^+e^- \to q\,\bar q$,
shown in Fig.~\ref{fig:FO-Tree}, identifying the thrust axis with the direction of the produced quark.
The structure of singular logarithms
and power corrections is completely identical to that of averaged event-shapes. Terms with any other angular
dependence cannot contain singular terms. We will explicitly verify this at $\mathcal{O}(\as)$ (analytically)
in Sec.~\ref{sec:LO} and at $\mathcal{O}(\as^2)$ (numerically) in Sec.~\ref{sec:NLO}.
The same result was found in \cite{Fleming:2007xt} for the case of doubly differential hemisphere
mass distribution.

We close this section by showing the explicit factorized form of the averaged event-shape distribution
\cite{Bauer:2008dt}
\begin{align}
\dfrac{1}{\sigma_0}\dfrac{\df \sigma_{\rm S}}{\df e} = H(Q,\mu)\!\int\! \df e_n\,\df e_{\bar n}\,\df e_s
J_n(e_n,\mu)\,J_{\bar n}(e_{\bar m},\mu)\, S(e_s,\mu)\,
\delta(e - e_s - e_n - e_{\bar n})\,.
\end{align}
The dominant non-perturbative effects are encoded in the soft function. The jet functions
describe collinear radiation (black arrows in Fig.~\ref{fig:thrust-axis}) and the soft function describes
large angle soft radiation (green wiggly lines in Fig.~\ref{fig:thrust-axis}).

\section{LO Distribution}
\label{sec:LO}

\begin{figure}[t!]
\begin{center}
\includegraphics[width=7cm]{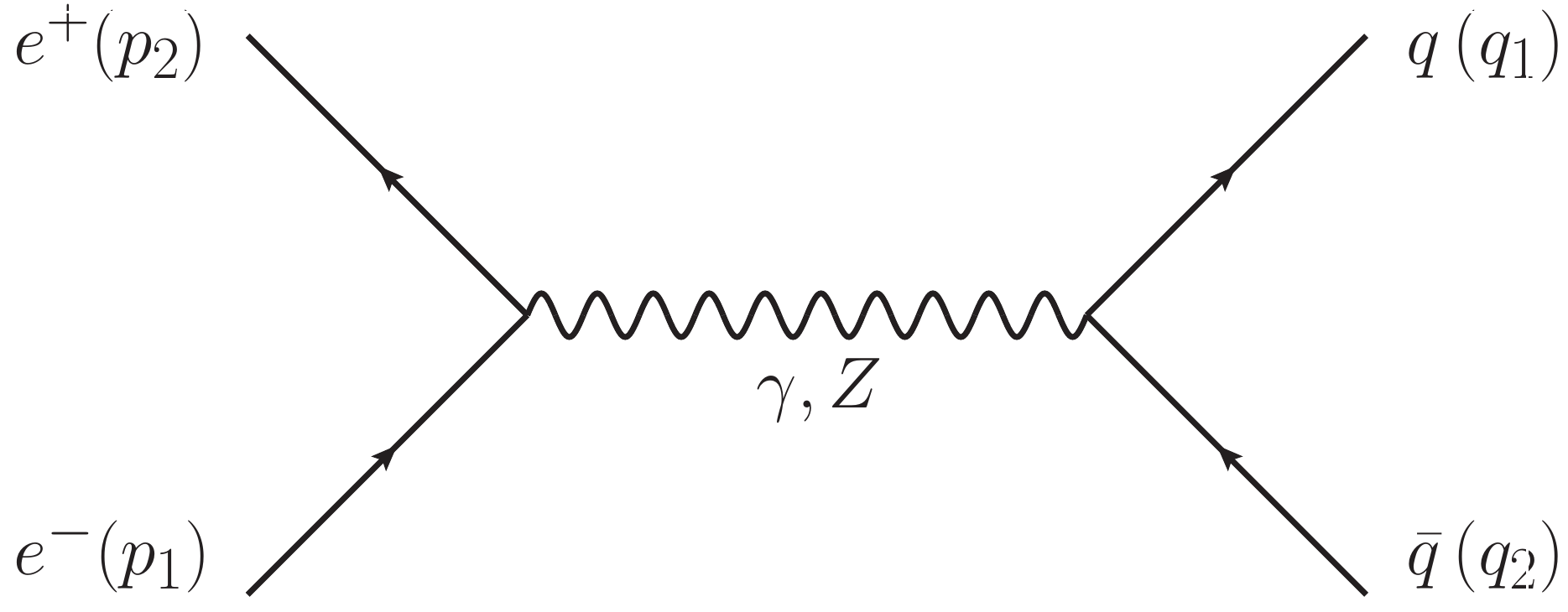}
\caption{The ${\cal O}(\as^0)$ contribution to oriented event-shapes.
\label{fig:FO-Tree}} 
\end{center}
\end{figure}

In this section we explicitly compute the oriented distribution at
$\mathcal{O}(\as^0)$ and $\mathcal{O}(\as)$ for any event-shape.
Let us start with the tree-level result, which we know is purely singular, and
hence from the main result of Sec.~\ref{sec:SCET} is expected to be proportional to
$1+\cos^2\theta_T$. The diagram to be computed is shown in 
Fig.~\ref{fig:FO-Tree}. In this case the thrust axis is obviously aligned with the
direction of the (anti-)quark, and the event-shape variable can only take its
lowest order value. For most event-shapes this value is simply $0$. Accordingly
one only needs to compute the differential cross-section in $\cos\theta_T$ and
multiply the result by $\delta(e)$. We find
\begin{align}
\dfrac{1}{\sigma_0}\dfrac{\df \sigma^0}{\df\! \cos\theta_T\,\df e} \,=\,
\dfrac{3}{8}\,(1 + \cos^2\theta_T)\,\delta(e)\,.
\end{align}

The first non-trivial computation appears at $\mathcal{O}(\as)$. It is nevertheless possible
to carry out analytically the projection to $\cos\theta_T$. The final projection to
any particular event-shape can be performed afterwards in the usual fashion. For
this exercise, as a first step we find it convenient to parametrize the phase-space
in terms of the energies of the quark $E_1$ and anti-quark $E_2$ and two out of
the three angles formed by the incoming electron and the quark ($\theta_1$),
the anti-quark ($\theta_2$) and the gluon ($\theta_3$). More specifically we define
\begin{align}
\label{customary}
& E_1 = \dfrac{Q x_1}{2}\,,\qquad E_2 = \dfrac{Q x_2}{2}\,, \\
& x_1 + x_2 + x_3 = 2\,, \nonumber \\
& x_1 \cos\theta_1 + x_2 \cos\theta_2 + x_3 \cos\theta_3 = 0\,,\nonumber
\end{align}
as it is customary. The second line in~(\ref{customary}) follows 
from energy conservation, the last line from three-momentum conservation in
the beam direction. Using these variables and adding the flux factor we
find\,\footnote{This result is obtained
when resolving the energy-conserving Dirac-delta function by integrating the azimuthal
angle $\varphi_j$, which is taken to be measured with respect to $\varphi_i$.}

\begin{figure}[t!]
\begin{center}
\includegraphics[width=15cm]{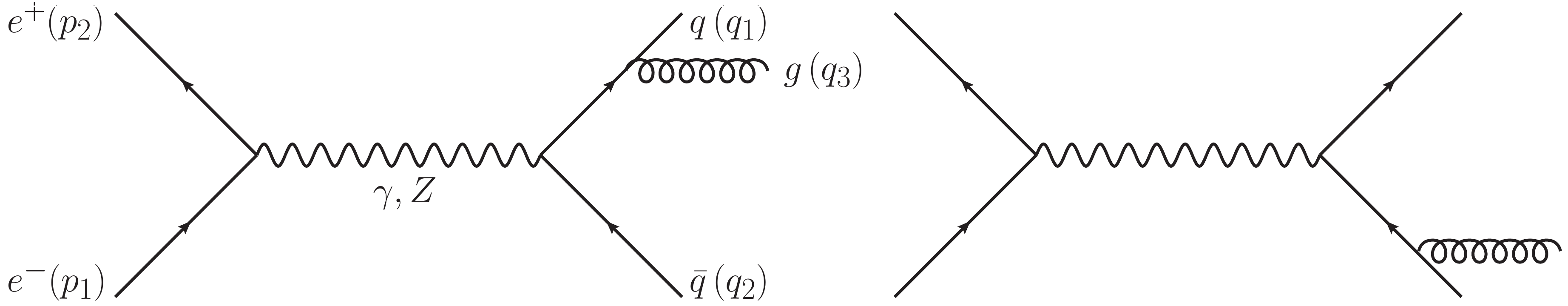}
\caption{Leading order corrections at ${\cal O}(\as)$ to oriented event-shapes.
\label{fig:FO-LO}} 
\end{center}
\end{figure}
\begin{figure}[t!]
\begin{center}
\includegraphics[width=4cm]{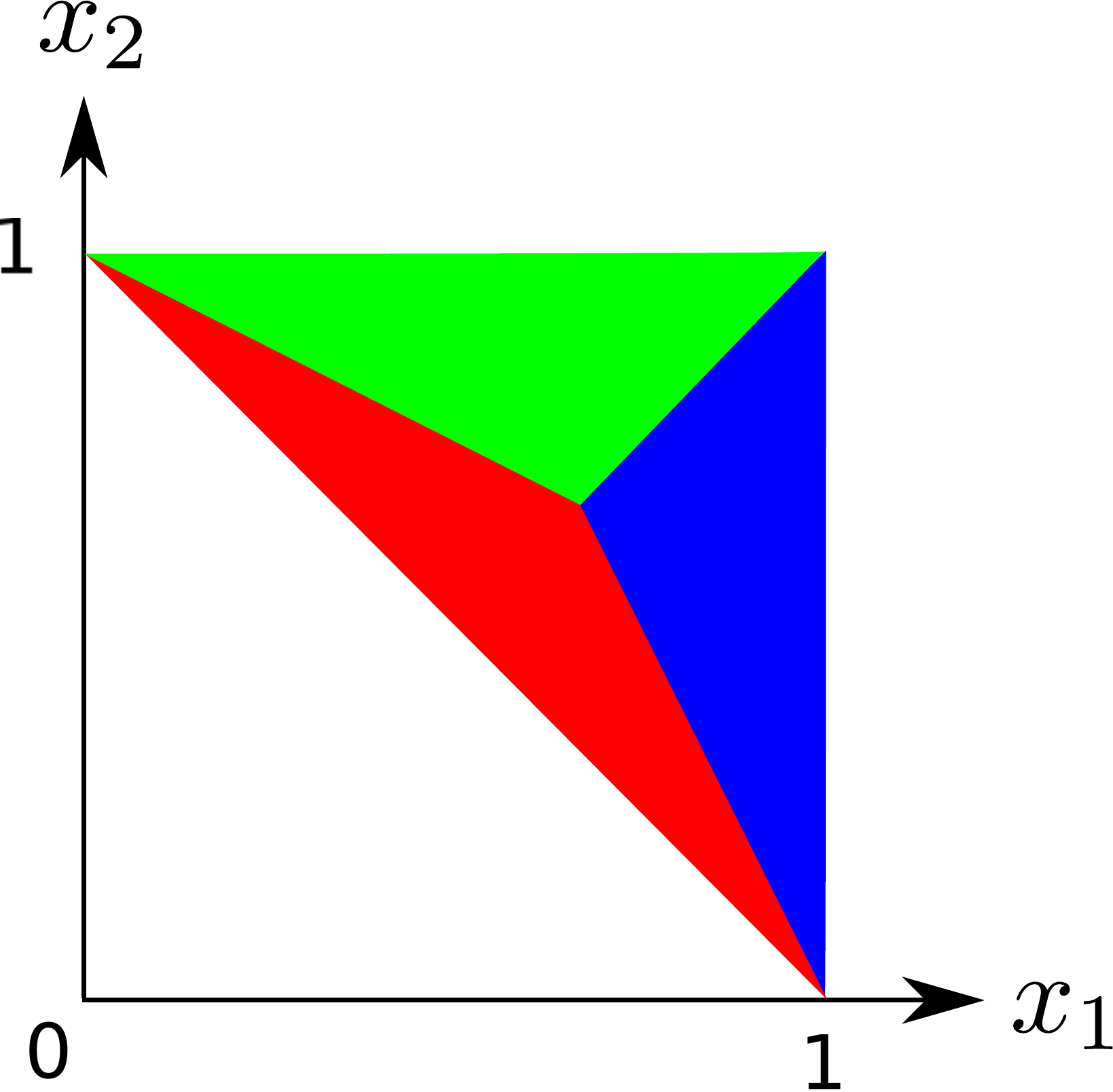}
\caption{Regions of phase-space with a common thrust axis. In the blue (red) region
the thrust axis points into the quark (anti-quark) three-momentum direction; in the green
region the thrust axis points into the gluon momentum.
\label{fig:thrust-projection}}
\end{center}
\end{figure}


\begin{align}
\dfrac{1}{2\,Q^2}\,\df \phi_3(P;q_1,q_2,q_3) &= \dfrac{1}{512\,\pi^3}\,\df x_1 \df x_2\,
\df\! \cos \theta_i\,\df \!\cos \theta_j\, 
\theta(\sin^2\theta_{ij})\, h_{ij}^{-1/2} \, \theta(h_{ij})~, \nonumber \\
\cos\theta_{ij} & = \dfrac{2x_k + x_i x_j-2}{x_i x_j}~, \nonumber \\
\sin\theta_{ij} & = \dfrac{2}{x_ix_j}\sqrt{(1-x_1)(1-x_2)(1-x_3)}~,\nonumber\\
h_{ij} &=\sin^2\theta_{ij}-\cos^2\theta_i-\cos^2\theta_j
+2\cos\theta_{ij}\cos\theta_i\cos\theta_j~, 
\end{align}
where $i\neq j \neq k$ take values $1$, $2$ or $3$. The function
$h_{ij}$ can conveniently be written as
\begin{align}
& h_{ij} = (\cos\theta_{ij}^+ - \cos\theta_i)(\cos\theta_i - \cos\theta_{ij}^-)~,\\
& \cos\theta_{ij}^{\pm} = \cos\theta_{ij}\cos\theta_j \,\pm\,
\sin\theta_{ij}\sin\theta_j = \cos(\theta_{ij} \mp \theta_j)~.\nonumber
\end{align}
The only integrals involving $h_{ij}$ that one needs contain powers of
$\cos^n\theta_i$ that can be trivially solved
\begin{align}
&\int\! \df\!\cos\theta_i\,\df\!\cos\theta_j\,h_{ij}^{-1/2} \theta(h_{ij})\cos^n\theta_i =\\
&\sqrt{\pi}\,\sum_{l=0}^n 
\binom{n}{l}\dfrac{(2\sin\theta_{ij}\sin\theta_j)^l\cos^{n-l}\theta_{ij}^+}{l!}\,
\Gamma\bigg(\dfrac{1}{2}+l\bigg)\,.\nonumber
\end{align}
Computing the diagrams depicted in Fig.~\ref{fig:FO-LO} we find
\begin{align}\label{eq:LO-angular-differential}
\dfrac{1}{\sigma_0}\dfrac{\df \sigma^{\rm LO}}
{\df x_1\df x_2 \df\!\cos\theta_i\df\!\cos\theta_j} =
\dfrac{3\as}{16\pi^2}\,C_F\,h_{ij}^{-1/2} \theta(h_{ij})\,
\dfrac{(1+\cos^2\theta_1)\,x_1^2 + (1+\cos^2\theta_2)\,x_2^2}
{(1-x_1)(1-x_2)}\,.
\end{align}
Integrating over $\cos\theta_1$ and $\cos\theta_2$ we recover the averaged result of Ref.~\cite{Ellis:1980nc}
\begin{align}\label{eq:LO-averaged}
 \dfrac{1}{\sigma_0}\dfrac{\df \sigma^{\rm LO}}{\df x_1\df x_2} =
 \dfrac{\as}{2\pi}\,C_F\,
 \dfrac{x_1^2 + x_2^2}
{(1-x_1)(1-x_2)}\,.
\end{align}
The next step is to project Eq.~(\ref{eq:LO-angular-differential}) onto $\cos\theta_T$. This is achieved with
the following projecting delta function:
\begin{align}
 \delta_T^{(3)} & = \theta(x_1 - x_2)\,\theta(x_1 - x_3)\, \delta(\cos\theta_T - \cos\theta_1)\\
 &+\, 
\theta(x_2 - x_1)\,\theta(x_2 - x_3)\, \delta(\cos\theta_T - \cos\theta_2) \nonumber\\
& + \,\theta(x_3 - x_1)\,\theta(x_3 - x_2)\, \delta(\cos\theta_T - \cos\theta_3)\,.\nonumber
\end{align}
The regions delimited by the various pairs of $\theta$ functions are shown in Fig.~\ref{fig:thrust-projection}.
Let us define the polar-angle phase-space integral that projects onto the thrust axis as
\begin{align}
\df \Phi_{ij}^T = \df\!\cos\theta_i\,\df\!\cos\theta_j\,h_{ij}^{-1/2} \theta(h_{ij})\, \delta_T^{(3)}~.
\end{align}

Then the projection of the three different angular-dependent pieces is straightforward:
\begin{align}\label{eq:project-pieces}
\dfrac{1}{\pi}\int\!  \df \Phi_{ij}^T = & \, 1\,, \\
\dfrac{1}{\pi}\int\! \df \Phi_{ij}^T \cos^2\theta_1 = & \,\cos^2\theta_T + \dfrac{1}{2}(1 - 3 \cos^2\theta_T)
\bigg[\theta(x_2 - x_1)\,\theta(x_2 - x_3) \sin^3\theta_{12} \nonumber\\
& +
\theta(x_3 - x_1)\,\theta(x_3 - x_2)\sin^3\theta_{13}\bigg]\,, \nonumber\\
\dfrac{1}{\pi}\int\! \df \Phi_{ij}^T \cos^2\theta_2 = & \, \cos^2\theta_T + \dfrac{1}{2}(1 - 3 \cos^2\theta_T)
\bigg[\theta(x_1 - x_2)\,\theta(x_1 - x_3) \sin^3\theta_{12} \nonumber\\
& +
\theta(x_3 - x_1)\,\theta(x_3 - x_2)\sin^3\theta_{23}\bigg]\,. \nonumber
\end{align}
Using the results of Eq.~(\ref{eq:project-pieces}) in (\ref{eq:LO-angular-differential}) we find
\begin{align}\label{eq:LO-projected}
& \dfrac{1}{\sigma_0}\dfrac{\df \sigma^{\rm LO}}{\df x_1\df x_2\,\df\!\cos\theta_T}  =
\dfrac{3}{8} \,(1 + \cos^2\theta_T)\, \dfrac{1}{\sigma_0}\dfrac{\df \sigma^{\rm LO}}{\df x_1\df x_2} +
\dfrac{3\as}{8\pi}\,C_F (1 - 3\cos^2\theta_T)\\
&\times (1 - x_3)\Bigg[\dfrac{\theta(x_1 - x_2)\,\theta(x_1 - x_3)}{x_1^2}+
\dfrac{\theta(x_2 - x_1)\,\theta(x_2 - x_3)}{x_2^2}
+\dfrac{2\,\theta(x_3 - x_1)\,\theta(x_3 - x_2)}{x_3^2}\Bigg]\,.
\nonumber
\end{align}
The second line in Eq.~(\ref{eq:LO-projected}) 
is in agreement with the results of Ref.~\cite{Lampe:1992au}.
Integrating over $\cos\theta_T$ one recovers the averaged result of Eq.~(\ref{eq:LO-averaged}).
From Eq.~(\ref{eq:LO-projected}) it is obvious that only the term proportional to $(1 + \cos^2\theta_T)$
can contain singular terms, as predicted by SCET in Eq.~(\ref{eq:final-SCET}). Finally, when projected
onto any event-shape $e$ one has the generic form
\begin{align}\label{eq:LO-projected-Event-Shape}
\dfrac{1}{\sigma_0}\dfrac{\df \sigma^{\rm LO}}{\df e\,\df\!\cos\theta_T}  =
\dfrac{3}{8} \,(1 + \cos^2\theta_T)\, \dfrac{1}{\sigma_0}\dfrac{\df \sigma^{\rm LO}}{\df e} +
(1 - 3\cos^2\theta_T) \,\dfrac{1}{\sigma_0}\dfrac{\df \sigma_{\rm ang}^{\rm LO}}{\df e}\,.
\end{align}
We shall demonstrate in Sec.~\ref{sec:Angular} that the structure of Eq.~(\ref{eq:LO-projected-Event-Shape})
holds to any order in perturbation theory and even beyond perturbative QCD.

It turns out that the term proportional to $(1-3\cos^2\theta_T)$ is much smaller
that the one proportional to $(1+\cos^2\theta_T)$. This is partially due to the
fact that the former is purely non-singular while the latter has singular terms
which numerically dominate. However, even if the singular terms are subtracted
the second structure is at most $20\%$ ($15\%$ on average) of the first one.
This behavior persists at higher orders.

\begin{figure*}[t!]
\subfigure[]{
\includegraphics[width=0.475\textwidth]{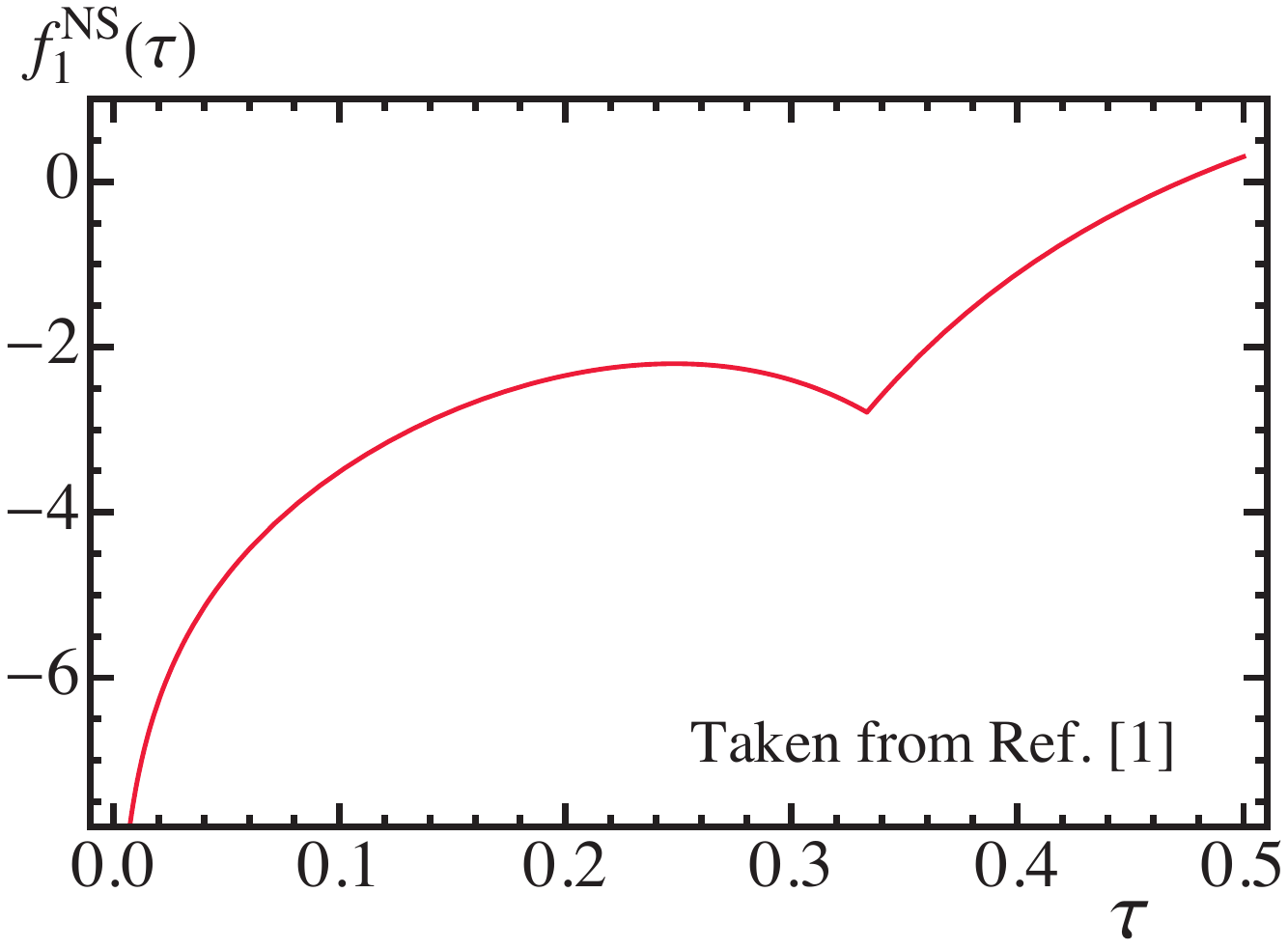}
\label{fig:thrust-NS-LO}
}
\subfigure[]{
\includegraphics[width=0.475\textwidth]{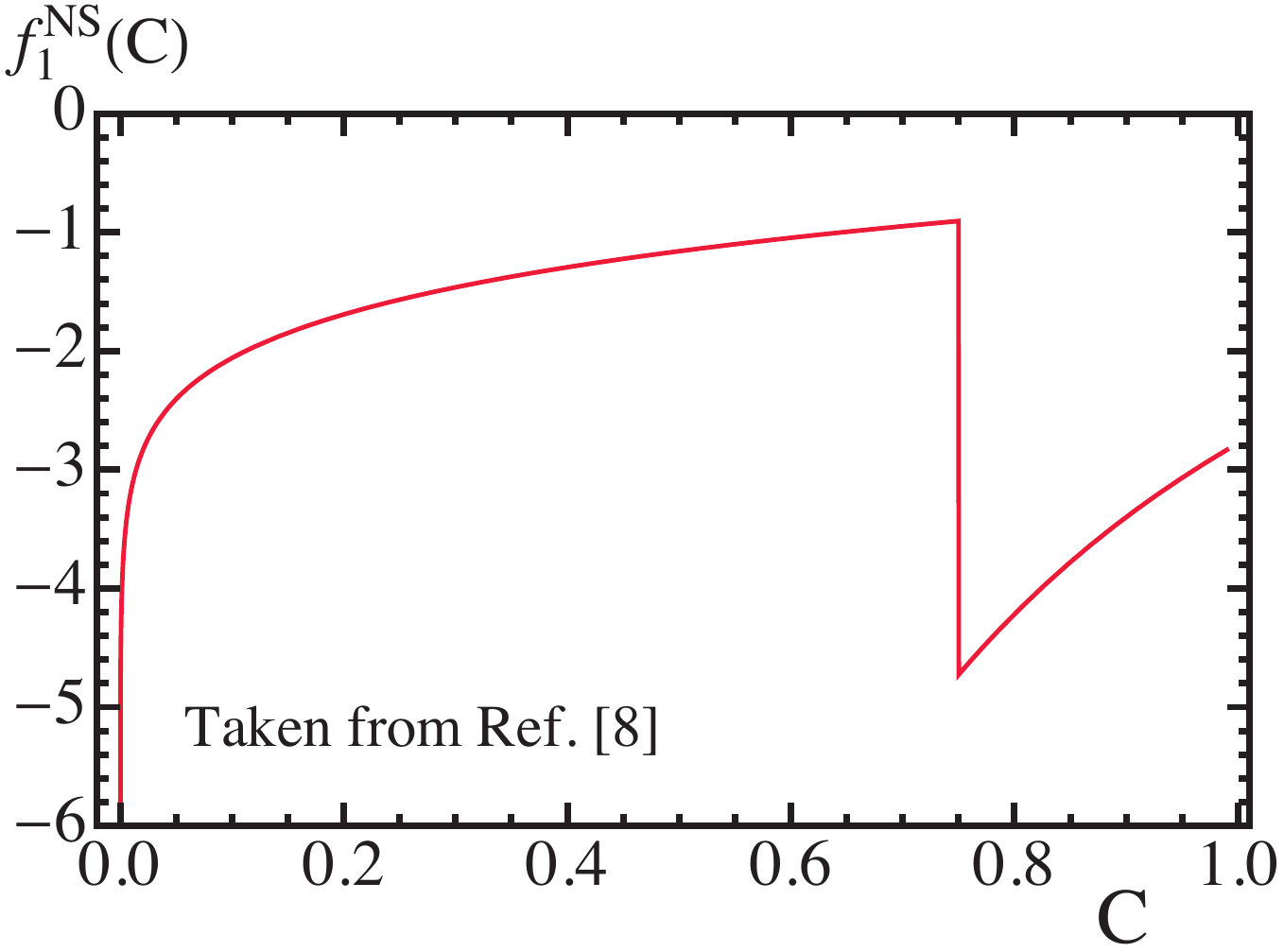}
\label{fig:Cparam-NS-LO}
}
\subfigure[]{
\includegraphics[width=0.475\textwidth]{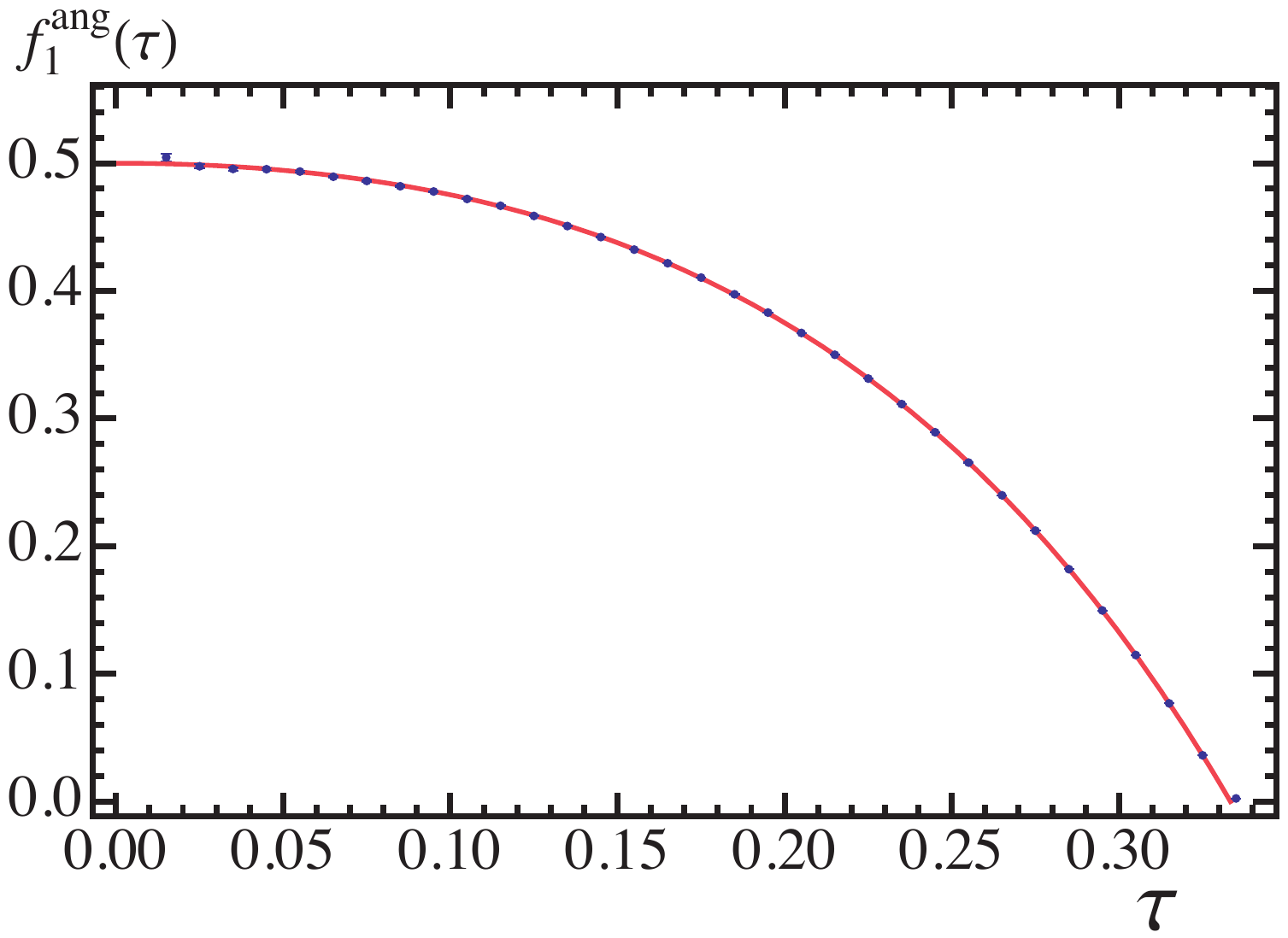}
\label{fig:thrust-Ang-LO}
}
\subfigure[]{
\includegraphics[width=0.475\textwidth]{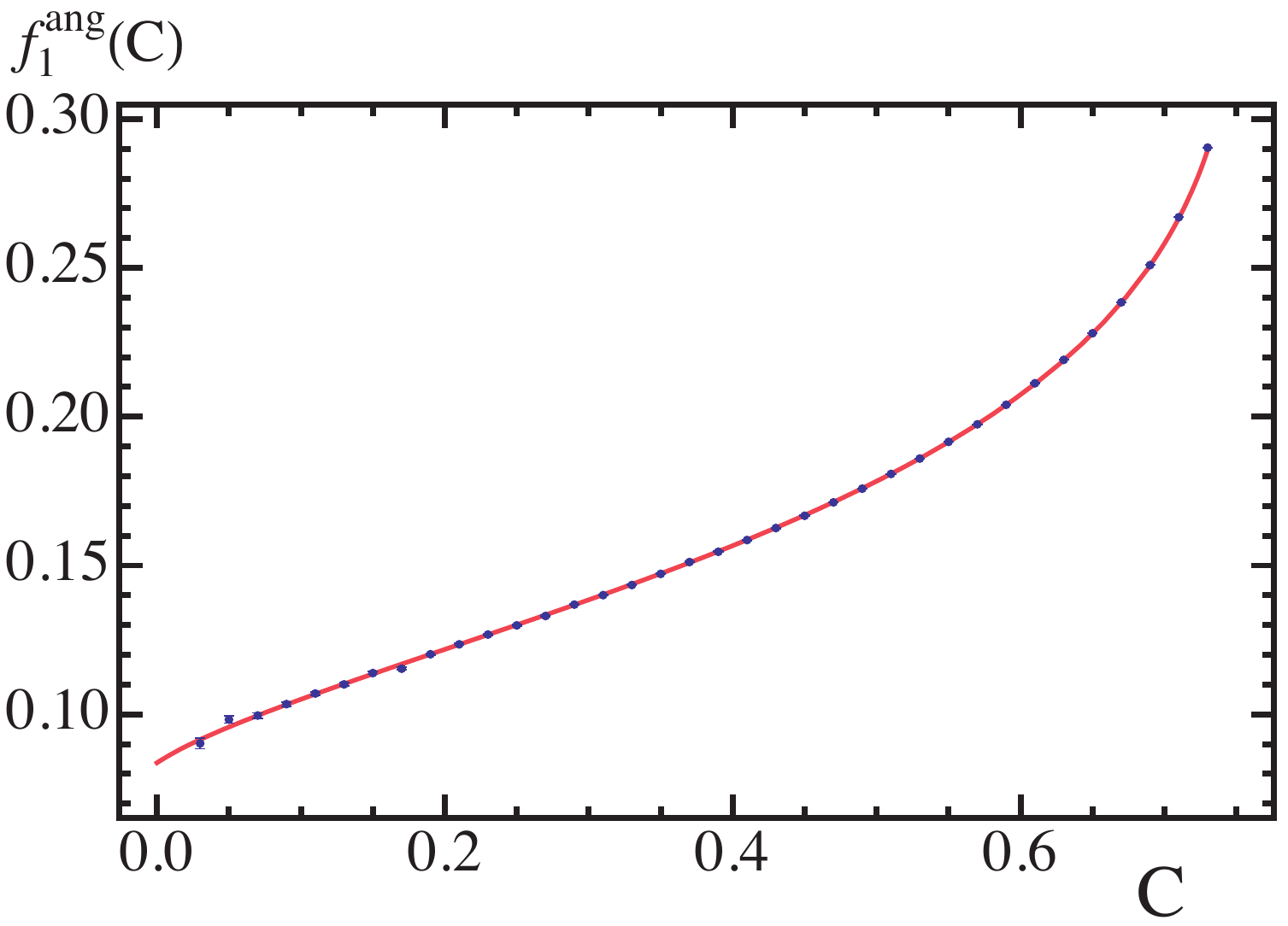}
\label{fig:Cparam-Ang-LO}
}
\vspace{-0.2cm}
\caption{Averaged non-singular and angular distributions at $\mathcal{O}(\as)$.
The left top panel (a) shows the thrust non-singular averaged distribution and the right
top panel (b) shows the $C$-parameter non-singular averaged distribution.
The left bottom panel (c) corresponds to the thrust angular distribution and
the right bottom panel (d) to the $C$-parameter angular distribution. The solid line
is obtained with an analytical computation, whereas blue dots with tiny error
bars are the LO output of our Event2 runs. The exact definitions of $f_1^{\rm NS}$
and $f_1^{\rm ang}$ can be found in Eq.~(\ref{eq:FO-expansion}).
The fact that the angular distributions tend to a finite value
for $\tau$ or $C$ tending to zero discards singular terms, as predicted by SCET.
\label{fig:LO-thrust-Cparam}}
\end{figure*}

As a final comment, one can calculate the total-oriented cross-section, completely inclusive in the final-state
hadrons but still differential in the thrust direction. It again follows the pattern of
Eq.~(\ref{eq:LO-projected-Event-Shape}):
\begin{align}\label{eq:angular-Rhad}
&\dfrac{1}{\sigma_0}\dfrac{\df \sigma}{\df\!\cos\theta_T} = \dfrac{3}{8} \,(1 + \cos^2\theta_T)\,R_{\rm had}
+(1 - 3\cos^2\theta_T) \, R_{\rm ang}\,,\\
&R_{\rm had} = 1 + \dfrac{\alpha_S}{\pi} \, \dfrac{3\,C_F}{4}
+ \mathcal{O}(\as^2)
=
1 + \dfrac{\alpha_S}{\pi}\,R_1^{\rm had} + \mathcal{O}(\as^2)\,,\nonumber \\
&R_{\rm ang} = \dfrac{\as}{\pi}\,\dfrac{3\,C_F}{8}\, 
\Big[\,8\,\log\Big(\dfrac{3}{2}\Big)-3\,\Big]
+\mathcal{O}(\as^2) =
\dfrac{\as}{\pi}\,R_1^{\rm ang} + \mathcal{O}(\as^2)\,,\nonumber
\end{align}
which has been obtained by integrating Eq.~(\ref{eq:LO-projected}).
Our result agrees with that of Ref.~\cite{Lampe:1992au}.
In Eq.~(\ref{eq:angular-Rhad}) 
we have denoted $R_{\rm had}$ as the averaged total cross-section
and $R_{\rm ang}$ as the angular total cross-section.
We will compute numerically
the $\mathcal{O}(\as^2)$ contributions in Sec.~\ref{sec:NLO}.
As a closing remark we compute the angular term for the thrust event-shape:
\begin{align}\label{eq:Thrust-Ang-LO}
\dfrac{1}{\sigma_0}\dfrac{\df \sigma_{\rm ang}^{\rm LO}}{\df \tau} =
\dfrac{3\,\as}{8\,\pi}\,C_F\,\dfrac{(1-3\,\tau)(1+\tau)}{(1-\tau)^2}
= \frac{\as}{\pi} \, f_1^{\rm ang} (\tau)\,,
\end{align}
result in agreement with Ref.~\cite{Lampe:1992au}.
For $C$-parameter the angular distribution can be expressed as the sum of two integrals which
can easily be integrated numerically
\begin{align}\label{eq:Cparam-Ang-LO}
&f_1^{\rm ang}(C) = \dfrac{C_F}{4}\Bigg[
\int_{s^-}^{{\tilde s}^-}\!\df s\,\dfrac{(s-1)^3}{(2-s)}\,
\dfrac{1}
{[1 + 2\,\tilde{c} - s\,(1+{\tilde c})]^{\frac{3}{2}}\sqrt{(2-s)(1+{\tilde c})(s-s_+)(s-s_-)}}\nonumber\\
&+\int_{\tilde{s}^-}^{s^+}\!\df s\,
\dfrac{2\,(2-s)(s-1)^3}{[1 + 2\,\tilde{c} - s\,(1+{\tilde c})]^{\frac{3}{2}}
\sqrt{(2-s)(1+{\tilde c})(s-s_+)(s-s_-)}}
\nonumber\\
&\qquad\times\dfrac{1}{\Bigg(s + \sqrt{\dfrac{(2-s)(1+{\tilde c})(s-s_+)(s-s_-) }
{1 + 2\,\tilde{c} - s\,(1+{\tilde c})}}\,\,\Bigg)^2}\,\,\Bigg]\,,\nonumber\\
& s^\pm = \dfrac{3\pm\sqrt{1-8\,\tilde{c}}}{2\,(1+\tilde{c})}\,,\qquad
{\tilde s}^\pm = 2 -\dfrac{1}{2}\,s^\mp\,,
\qquad \tilde{c} = \dfrac{C}{6}\,.
\end{align}
$f_1^{\rm ang}$ for thrust and $C$-parameter are shown, along with the corresponding non-singular
distributions, in Fig.~\ref{fig:LO-thrust-Cparam}\,\footnote{The non-singular distribution is defined
as the full fixed-order distribution with the singular terms subtracted out.
The singular terms are obtained from the SCET factorization theorem when no
resummation is carried out, that is with all renormalization scales set equal. For thrust
the singular terms can be found in \cite{Becher:2008cf}, for Heavy-Jet Mass in \cite{Chien:2010kc}
and for $C$-parameter in \cite{Hoang:2013}.}.

\section{Angular Dependence to All Orders}
\label{sec:Angular}
As long as electroweak interactions between the initial and the final state are ignored,
at any order (or equivalently for an arbitrary number $n$ of final-state particles) the matrix
element is given by the contraction of a leptonic and
a hadronic tensor. The leptonic tensor has the same form as in Eq.~(\ref{eq:lepton-tensor}), and
the hadronic tensor has the general form~\footnote{In this section we consider
the general case in which final-state particles could have a mass.}
\begin{align}
H^{\mu\nu} & = A_0(q_r\cdot q_s,m_h)\, g^{\mu\nu} +
\sum_{i,j=1}^n A_{ij}(q_r\cdot q_s,m_h)\, (q_i^\mu q_j^\nu + q_i^\nu q_j^\mu) \nonumber\\
& +
\sum_{i,j,k,l=1}^n A_{ijkl}(q_r\cdot q_s,m_h) (q_i^\mu q_{j\alpha}q_{k\beta}q_{l\gamma}
\varepsilon^{\alpha\beta\gamma\nu} + q_i^\nu q_{j\alpha}q_{k\beta}q_{l\gamma}
\varepsilon^{\alpha\beta\gamma\mu})\nonumber \\
& + \sum_{i,j,k=1}^n A_{ijk}(q_r\cdot q_s,m_h)\, q_{i\alpha}\,q_{j\beta}\,q_{k\gamma}\,
\varepsilon^{\alpha\beta\gamma\mu}\,
q_{i\delta}\,q_{j\rho}\,q_{k\kappa}\,
\varepsilon^{\delta\rho\kappa\nu}~,
\end{align}
where we have used the fact that the hadronic tensor is symmetric when interchanging the indices
$\mu$ and $\nu$ (Imaginary parts generated by virtual corrections produce imaginary, antisymmetric
tensors, which cancel upon contraction with the leptonic tensor. The symmetric part is real).
When contracting leptonic and hadronic tensors one obtains a piece which is
independent of the direction of the incoming electron and positron, and a piece which depends
of the orientation of the hadrons with respect to the beam. We get
\begin{align}\label{eq:general-contraction}
L_{\mu\nu}H^{\mu\nu} & = A_{0}(q_r\cdot q_s,m_h) +
\sum_{i,j=1}^n 2\,A_{ij}(q_r\cdot q_s,m_h) 
[(p_1\cdot q_i)(p_2\cdot q_j) + (p_2\cdot q_i)(p_1\cdot q_j)] \nonumber \\
& +
\sum_{i,j,k,l=1}^n A_{ijkl}(q_r\cdot q_s,m_h) [(p_1\cdot q_i)\, q_{j\alpha}\,q_{k\beta}\,q_{l\gamma}\,p_{2\mu}
\varepsilon^{\alpha\beta\gamma\mu} + (p_2\cdot q_i)\, q_{j\alpha}\,q_{k\beta}\,q_{l\gamma}\,p_{1\mu}
\varepsilon^{\alpha\beta\gamma\mu}]\nonumber\\
&+ \sum_{i,j,k=1}^n A_{ijk}(q_r\cdot q_s,m_h)\, q_{i\alpha}\,q_{j\beta}\,q_{k\gamma}\,p_{1\mu}\,
\varepsilon^{\alpha\beta\gamma\mu}
q_{i\delta}\,q_{j\rho}\,q_{k\kappa}\,p_{2\nu}\,
\varepsilon^{\delta\rho\kappa\nu}~.
\end{align}
We are interested in the dependence on the angles of the final-state particles
with the incoming electron. It comes solely from scalar products of the type
\begin{align}\label{eq:mass-angle}
& (p_1\cdot q_i)(p_2\cdot q_j) + (p_2\cdot q_i)(p_1\cdot q_j) =
\dfrac{Q^2}{2}\,E_i\,E_j - 2\,(\vec{p}_1\cdot\vec{q}_i) (\vec{p}_1\cdot\vec{q}_j)~,\\
& q_{i\alpha}\,q_{j\beta}\,q_{k\gamma}\,p_{1\mu}\,
\varepsilon^{\alpha\beta\gamma\mu}
q_{i\delta}\,q_{j\rho}\,q_{k\kappa}\,p_{2\nu}\,
\varepsilon^{\delta\rho\kappa\nu} = \dfrac{Q^2}{4}\big[ \vec{q}_i\cdot (\vec{q}_j\times \vec{q}_k)\big]^2
\nonumber\\
& \qquad\qquad\qquad\qquad - \Big[\vec{p}_1\cdot(E_i\, \vec{q}_j\times \vec{q}_k + E_j\, \vec{q}_k\times \vec{q}_i
+ E_k\, \vec{q}_i\times \vec{q}_j)\Big]^2~,\nonumber\\
&(p_1\cdot q_i)\, q_{j\alpha}\,q_{k\beta}\,q_{l\gamma}\,p_{2\mu}\,
\varepsilon^{\alpha\beta\gamma\mu}  + (p_2\cdot q_i)\, q_{j\alpha}\,q_{k\beta}\,q_{l\gamma}\,p_{1\mu}\,
\varepsilon^{\alpha\beta\gamma\mu} = \dfrac{Q^2}{2}\, \vec{q}_j\cdot (\vec{q}_l\times \vec{q}_k)
\nonumber\\
& \qquad\qquad\qquad\qquad - 
2\,(\vec{p}_1\cdot \vec{q}_i)\big[\,\vec{p}_1\cdot(E_j\, \vec{q}_k\times \vec{q}_l + E_k\, \vec{q}_l\times \vec{q}_j
+ E_l\, \vec{q}_j\times \vec{q}_k)\,\big]~.\nonumber
\end{align}
The expression above is invariant under the transformation $\vec{p}_1\to -\,\vec{p}_1$.
Plugging Eq.~(\ref{eq:mass-angle}) into Eq.~(\ref{eq:general-contraction}) one gets
\begin{align}\label{eq:general-angular}
L_{\mu\nu}H^{\mu\nu} &= C_0(q_k\cdot q_l,m_h) + 
\sum_{i,j=1}^n C_{ij}(q_r\cdot q_s,m_h)\, (\vec{p}_1\cdot \vec{q}_i)(\vec{p}_1\cdot \vec{q}_j) \\
& +
\sum_{i,j,k} C_{ijk}(q_r\cdot q_s,m_h)\, (\vec{p}_1\cdot \vec{q}_i) \,
[\,\vec{p}_1\cdot(\vec{q}_j\times \vec{q}_k)\,]\nonumber\\
& +
\sum_{i,j,k,l} C_{ijkl}(q_r\cdot q_s,m_h)\,
[\,\vec{p}_1\cdot(\vec{q}_i\times \vec{q}_j)\,] \,
[\,\vec{p}_1\cdot(\vec{q}_k\times \vec{q}_l)\,]\,.
\nonumber
\end{align}
The next step is to integrate Eq.~(\ref{eq:general-angular}) with the phase-space
of $n$ particles and a delta function that projects out $\cos\theta_T$.

In order to proof that the only angular structures that one can find at any order
are precisely those found at LO we will make extensive use of two important 
properties of the thrust axis. From the definition of thrust $T$ one has
\begin{align}
\Big(\sum_i|\vec{q}_i|\Big)\,T & = \max_{\hat{n}}\sum_i |\hat{n}\cdot \vec{q}_i| =
\max_{\hat{n}}\bigg|\hat{n}\cdot
\Big(\sum_{i\in a}\vec{q}_i\,-\sum_{i\in b}\vec{q}_i\Big)\bigg|\\
& = 2\max_{\hat{n}}\bigg|\hat{n}\cdot\sum_{i\in a}\vec{q}_i\bigg|
\equiv 2\max_{\hat{n}}\big|\hat{n}\cdot\vec{P}_a\big|\,.\nonumber
\end{align}
Where the hemisphere $a$ contains all particles $\vec{q}_i$ satisfying
$\hat{n}\cdot \vec{q}_i>0$, whereas hemisphere $b$ has particles with
$\hat{n}\cdot \vec{q}_i<0$. In the one-to-last last equality we have used
three-momentum conservation. In the last equality we have defined the total
four-momentum in one hemisphere
\begin{align}
P^\mu_a = \sum_{i\in a}q^\mu_i\,,\qquad P^\mu_b = \sum_{i\in b}q^\mu_i\,, 
\end{align}
and of course $\vec{P}_a = -\,\vec{P}_b$. We use the notation $P_{a,b}^2 \equiv S_{a,b}$,
which are the masses of each hemisphere. The fist important property of the thrust
axis is that particles can be clustered together (that is they can substituted by
a pseudo-particle whose momentum is the sum of the momenta of the clustered 
particles) as long as they belong to the same hemisphere. The next step is to find
the unitary vector $\hat n$ which maximizes $T$. It is obvious that this is achieved
if $\hat n$ points in the direction of $\vec{P}_a$ itself
\begin{align}
\hat{n} = \dfrac{\vec{P}_a}{|\vec{P}_a|}\,,\qquad
T = \dfrac{2\,|\vec{P}_a|}{\sum_i|\vec{q}_i|}\,.
\end{align}
Hence $T$ can be thought as the length of the longest possible vector that can be 
formed by clustering particles together, normalized to half of the sum of the
magnitudes of the final-state three-momenta.

The phase-space of $n$ particles is
\begin{align}
\df \phi_n(P;\,q_1,\ldots,q_n) & =
\Bigg(\prod_{i=1}^n\,\dfrac{\df ^3\vec{q}_i}{2E_i(2\pi)^3}\Bigg)
(2\pi)^4\,\delta^{(4)}\Big(P - \sum_{i=1}^n q_i\Big)\,,\\
E_i& = \sqrt{\vec{q}_i^{\,2} + m_i^2}\,,\nonumber\\
\df ^3\vec{q}_i & = \vec{q}_i^{\,2}\,\df |\vec{q}_i|\,
\df \varphi_i\,\df \!\cos\theta_i\,.\nonumber
\end{align}
Here $P$ represents the total momentum of the incoming leptons in the 
center-of-mass frame: $P = ( Q, \,\vec{0} )$. The phase-space 
integral can be decomposed in a series of sectors such that the $n$ particles can
be clustered into two hemispheres containing $k$ and $n-k$ particles, respectively.
This decomposition can be implemented by suitable theta functions which add up to one.
For the case of three partons worked out in Sec.~\ref{sec:LO} these $\theta$'s read
\begin{equation}
\theta(x_1 - x_2)\,\theta(x_1 - x_3) \,+\,
\theta(x_2 - x_1)\,\theta(x_2 - x_3) \, + \,
\theta(x_3 - x_1)\,\theta(x_3 - x_2) = 1\,.
\end{equation}
For each one of these sectors the phase-space factorizes in the following
way:
\begin{align}\label{eq:phase-space-factor}
\!\!\!\!\!\!\df \phi_n(P; q_1, \ldots, q_n) =\!
\int\!\dfrac{\df S_a}{2\pi}\dfrac{\df S_b}{2\pi}\,
\df \phi_2(P; P_a, P_b)\,\df \phi_k(P_a; q_1, \ldots, q_k)\,
\df \phi_{n-k}(P_b; q_{k+1}, \ldots, q_n)\,.
\end{align}
The projecting delta function in this sector reads
\begin{align}\label{eq:general-projecting}
\delta_T = \delta \Bigg(\!\cos\theta_T - 
\dfrac{\sum_{i=1}^k |\vec{q}_i| \cos\theta_{i}}
{|\sum_{i=1}^k \vec{q}_i|}\Bigg).
\end{align}
In Eq.~(\ref{eq:phase-space-factor}), the initial momentum in $\df \phi_n$
is used in the center-of-mass frame,
$\vec{P}=\vec{0}$ but in $\df \phi_k$ and $\df \phi_{n-k}$ is not, 
$\vec{P}_{a,b}\neq\vec{0}$. Moreover, the azimuthal $\phi$ and polar $\theta$
angles in $\df \phi_n$ are measured with respect to the direction of the beam
(that is with respect to the incoming electron), whereas for $\df \phi_k$ and
$\df \phi_{n-k}$ they are measured with respect to $\vec{P}_a$. One should integrate
a minimal amount of variables such that one can still later project onto any observable.
We shall see that one only needs to integrate over one trivial azimuthal angular dependence
to show that the pattern of Eq.~(\ref{eq:LO-projected-Event-Shape}) prevails to all orders.

Next, we integrate $\df^3\vec{P}_b$, $\df^3\vec{p}_1$ and
$\df ^3\vec{p}_{k+1}$ with the three corresponding spatial delta functions.
This simplifies Eq.~(\ref{eq:general-projecting}) to $\delta_T = \delta(\cos\theta_T - \cos\theta_a)$
and the two-particle phase-space can be resolved completely
\begin{align}
\df \phi_2(P; P_a, P_b) = \dfrac{1}{16\pi}\dfrac{\lambda^{\frac{1}{2}}(Q^2,S_a,S_b)}{Q^2}
\,\df\! \cos\theta_a\,,
\end{align}
where $\lambda(a,b,c) = (a + b - c)^2- 4\,a\,b$ stands for the completely symmetric K\"allen function.
In the matrix element Eq.~(\ref{eq:general-angular}) one has to make the replacements\,\footnote{In
principle one also has to make the replacement
$q_i\cdot q_j\to E_i E_j - |\vec{q}_i||\vec{q}_j|(\sin\theta_i\sin\theta_j\cos\varphi_{ij} + 
\cos\theta_i\cos\theta_j)$, but this does not affect the discussion on the dependence on $\cos\theta_T$.}
\begin{align}
&\vec{q}_1 = \vec{P}_a - \sum_{i=2}^k \vec{q}_i\,,\\
&\vec{q}_{k+1} = -\,\vec{P}_a - \sum_{i=k+2}^n \vec{q}_i\,, \nonumber\\
&\vec{p}_1\cdot\vec{P}_a  = \frac{Q}{2}|\vec{P}_a| \cos\theta_T\,,\nonumber\\
&\vec{p}_1\cdot \vec{q}_i = \frac{Q}{2}|\vec{q}_i|\,
(\sin\theta_T \sin\theta_i \cos\varphi_i + \cos\theta_T \cos \theta_i)\,,\nonumber\\
&\vec{p}_1\cdot(\vec{P}_a\times \vec{q}_i)  = 
-\,\dfrac{Q}{2}|\vec{P}_a||\vec{q}_i|\sin\theta_T\sin\theta_i\sin\varphi_i\,,\nonumber
\\
&\vec{p}_1\cdot(\vec{q}_i\times \vec{q}_j) = \dfrac{Q}{2}|\vec{q}_i||\vec{q}_j|
[\sin\theta_T (\sin\theta_i\cos\theta_j\sin\varphi_i - \sin\theta_j\cos\theta_i\sin\varphi_j)\nonumber
\\
& \qquad \qquad \qquad + \cos\theta_T \sin\theta_i\sin\theta_j\sin\varphi_{ji} ]  \,,\nonumber
\end{align}
with $\varphi_{ji} = \varphi_{j} - \varphi_{i}$.
Here we have taken the beam axis to lie on the $x-z$ plane, hence $\varphi_T = 0$. An important observation
is that $\sin\theta_T$ is always multiplied by a single power of $\cos\varphi_i$ or $\sin\varphi_i$.
Now Eq.~(\ref{eq:general-angular}) becomes
\begin{align}\label{eq:LHangular}
& L_{\mu\nu}H^{\mu\nu} = D^{(0)}(q_k\cdot q_l,m_h, \cos\varphi_i,\cos\theta_j) + 
\cos^2\theta_T \, D^{(1)}(q_k\cdot q_l,m_h,\cos\varphi_s,\sin\varphi_t,\cos\theta_m) \nonumber\\
& +
\sin\theta_T\cos\theta_T\Bigg[\sum_i\cos\varphi_i \, D^{(2)}_i(q_k\cdot q_l,m_h,\cos\theta_m)+
\sum_i\sin\varphi_i \, D^{(3)}_i(q_k\cdot q_l,m_h,\cos\theta_m)\nonumber\\
& +
\sum_{ijk}\cos\varphi_i \sin\varphi_{jk} \, D^{(4)}_i(q_k\cdot q_l,m_h,\cos\theta_m)
+
\sum_{ijk}\sin\varphi_i \sin\varphi_{jk} \, D^{(5)}_i(q_k\cdot q_l,m_h,\cos\theta_m)\Bigg]
\,,
\end{align}
where we have substituted $\sin^2\theta_T = 1 - \cos^2\theta_T$.
Additionally, in $\df \phi_k$ and $\df \phi_{n-k}$ one has to make the following replacements:
\begin{align}\label{eq:general-thetaT}
\vec{q}_1^{\,2} &= \vec{P}_a^{\,2} - 2\,|\vec{P}_a|\sum_{i=2}^k |\vec{q}_i|\cos\theta_i 
+ \sum_{i,j=2}^k |\vec{q}_i| |\vec{q}_j|(\sin\theta_i\sin\theta_j\cos\varphi_{ij} +
\cos\theta_i\cos\theta_j)\,, \\
\vec{q}_{k+1}^{\,2} & = \vec{P}_a^{\,2} + 2\,|\vec{P}_a|\sum_{i=k+2}^n |\vec{q}_i|\cos\theta_i
+ \sum_{i,j=k+1}^n |\vec{q}_i| |\vec{q}_j|(\sin\theta_i\sin\theta_j\cos\varphi_{ij} + 
\cos\theta_i\cos\theta_j)\,.\nonumber
\end{align}
The important observation is that the dependence on the azimuthal
angles is always through the difference of two angles.
Hence one can make the following change of variables: $\varphi_i \to \varphi_i + \varphi_k$ and completely
eliminate the dependence on $\varphi_k$ from the phase-space, which can then trivially be integrated. One should
choose $\varphi_k$ term by term in Eq.~(\ref{eq:LHangular}) such that the term proportional to
$\sin\theta_T\cos\theta_T$ vanishes upon azimuthal angular integration. This works trivially for
the terms with a single power of $\sin\varphi_i$ or $\cos\varphi_i$ by taking  $\varphi_k = \varphi_i$,
but also with terms such as $\cos\varphi_i \sin\varphi_{jl}$ or $\sin\varphi_i \sin\varphi_{jl}$.
In the latter case, if $i\neq j \neq l$ one can make $\varphi_k = \varphi_i$,
if $j = l$ then the term is automatically zero, and if $i = j$ then one can choose $\varphi_k = \varphi_l$. It is important to stress that no event-shape variable can depend on the overall
azimuthal angle around the thrust axis, and hence this integral can always be performed completely.
In a way we are averaging the event shape distribution with respect to the azimuthal direction
around the thrust axis.
Similarly, the phase space decomposition does not depend on a global azimuthal orientation either.
One would get sensitivity to it only if the beam is polarized. This global azimuthal angle is
represented in Fig.~\ref{fig:thrust-axis} as $\varphi_T$.

This procedure can be repeated sector by sector in exactly the same way. The outcome is that only
$\theta_T$-independent structures and terms proportional to $\cos^2\theta_T$ can arise. These can be conveniently
recast into $F_0(\cos\theta_T) = 3/8\,(1+\cos^2\theta_T)$ and $F_1(\cos\theta_T) = 1 - 3\cos^2\theta_T$. One can
afterwards project onto any event-shape $e$ by the appropriate delta function, and in full generality the result
can be expressed as
\begin{align}\label{eq:final-general}
\dfrac{1}{\sigma_0}\dfrac{\df \sigma}{\df \!\cos\theta_T\, \df e}
& = \dfrac{1}{\sigma_0}\dfrac{\df \sigma}{\df e}\, F_0(\cos\theta_T) + 
\dfrac{1}{\sigma_0}\dfrac{\df \sigma_{\rm ang}}{\df e}\, F_1(\cos\theta_T)\,,\\
\dfrac{1}{\sigma_0}\dfrac{\df \sigma}{\df e} & = \int_{-1}^1 \df\! \cos\theta_T \,
\dfrac{1}{\sigma_0}\dfrac{\df \sigma}{\df \!\cos\theta_T\, \df e}\,,\nonumber\\
\dfrac{1}{\sigma_0}\dfrac{\df \sigma_{\rm ang}}{\df e} & = \dfrac{3}{8}\int_{-1}^1 \df\! \cos\theta_T \,
(2-5\cos^2\theta_T)\,\dfrac{1}{\sigma_0}\dfrac{\df \sigma}{\df\! \cos\theta_T\, \df e}\,.\nonumber
\end{align}
We will refer to the cross-section multiplying $F_0$ as the ``averaged distribution'', and will refer
to the cross-section multiplying $F_1$ as the ``angular distribution''.
As a final comment we would like to emphasize that the proof is valid for partons (massless or massive)
as well as for hadrons, and hence the general structure of Eq.~(\ref{eq:final-general}) ``survives''
hadronization. In Refs.~\cite{Lampe:1992au,Hagiwara:2010cd,Abbiendi:1998at} the oriented distributions
are written in terms of transverse ($\sigma_T$) and longitudinal ($\sigma_L$) distributions. They are related
to our notation in a simple manner: $\sigma = \sigma_T + \sigma_L$ and $\sigma_{\rm ang} = 3/8\,\sigma_L$.

In Ref.~\cite{Lampe:1992au} it is shown that for three-parton processes the only two possible
structures are the same as in Eq.~(\ref{eq:final-general}). For processes with higher multiplicity the
longitudinal cross section is defined as the contraction of the hadronic tensor with the two longitudinal
polarization vectors of an intermediate massive vector boson, but no proof is given that this contraction
renders the angular structure predicted in Eq.~(\ref{eq:final-general}). Similarly Ref.~\cite{Abbiendi:1998at}
claims that Eq.~(\ref{eq:final-general}) is the most general result, without proof or quote. Finally, in
Ref.~\cite{Hagiwara:2010cd} an alternative proof for Eq.~(\ref{eq:final-general}) is presented. The proof relies
on current conservation (hence it would not apply to production of heavy quarks by an axial current,
and hence is not as general as ours).
Moreover, their demonstration assumes a particular Lorentz structure for the thrust-oriented hadronic tensor
in terms of four-vectors constructed out of the thrust axis direction. This structure is presented without a
rigorous proof, and it would certainly not hold for some other choices of the oriented axis. In a sense, our
demonstration could be used to proof their proposed hadronic tensor structure.

We finish this section with a discussion of the case in which parity-violating terms can arise.
The generalization is straightforward and it only requires to assign a direction to the
thrust axis\,\footnote{If one
is producing top-antitop pairs, one could for instance choose the thrust axis to point into the hemisphere
which contains the top.}. We will assume that the thrust axis points into he $\vec{P}_a$ direction.
Parity violating terms and virtual effects induce antisymmetric lepton and hadron tensors,
which are purely imaginary. In the lepton case one has antisymmetric terms though axial contributions
\begin{align}
L_{A}^{\mu \nu} \propto i\, p_{1\alpha}\, p_{2\beta}\, \varepsilon^{\alpha\beta\mu\nu}\,.
\end{align}
Similarly the antisymmetric part of the hadronic tensor would look like
\begin{align}
H_A^{\mu\nu} = \sum_{i,j}\bigg[A_{ij}\,q_{i\alpha}\, q_{j\beta}\, \varepsilon^{\alpha\beta\mu\nu} +
B_{ij}(q_i^\mu\,q_j^\nu - q_i^\nu\,q_j^\mu)\bigg]\,,
\end{align}
when contracting the antisymmetric parts of the leptonic and hadronic tensors one is left with two kind of
terms:
\begin{align}
L_{\mu\nu, A} \, H_A^{\mu\nu} \propto Q \sum_{i,j} \bigg[A_{ij}\,
\vec{p}_1\cdot(E_j\,\vec{q}_i - E_i\,\vec{q}_j)
+ 2\,B_{ij}\,\vec{p}_1\cdot(\vec{q}_i\times \vec{q}_j)\bigg]\,.
\end{align}
Following the same steps as for the parity-conserving terms one arrives to the conclusion that only
terms linear in $\cos\theta_T$ can arise (that is there are no terms proportional to $\sin\theta_T$,
since they cancel upon azimuthal angle integration). These terms contain singular and non-singular
terms, and the former can be treated in SCET in the same way as the averaged distribution.
\section{NLO Distribution}
\label{sec:NLO}
In this section we present results for
the non-singular and angular cross-sections for thrust, \mbox{$C$-parameter}, Heavy-Jet Mass and 
the sum of Hemisphere Masses~\footnote{The non-singular contributions to thrust, \mbox{$C$-parameter}
and Heavy-Jet Mass has been extracted from Refs.~\cite{Abbate:2010xh,Hoang:2013a,Hoang:2013}, respectively.
We collect these results here for completeness.}.
To that end we run the FORTRAN program Event2~\cite{Catani:1996jh,Catani:1996vz} with $6\times 10^{11}$ events.

The event-shape cross-section can be expanded in powers of $\as$.
Since we carry out resummation for the most singular terms of the averaged distribution, we do not use the
fixed-order counting for them, instead we use resummed counting (LL, NLL, 
etc\,\ldots). For the non-singular averaged distribution and the angular distribution 
we use the following expansion\,:
\begin{align}\label{eq:FO-expansion}
\dfrac{1}{\sigma_0}\dfrac{\df \sigma_{\rm NS}}{\df e} & = \dfrac{\as(Q)}{\pi}f_1^{\rm NS}(e)+
\bigg(\dfrac{\as(Q)}{\pi}\bigg)^2f_2^{\rm NS}(e)+
\bigg(\dfrac{\as(Q)}{\pi}\bigg)^3f_3^{\rm NS}(e)+\ldots\\
\dfrac{1}{\sigma_0}\dfrac{\df \sigma_{\rm ang}}{\df e} & = \dfrac{\as(Q)}{\pi}f_1^{\rm ang}(e)+
\bigg(\dfrac{\as(Q)}{\pi}\bigg)^2f_2^{\rm ang}(e)+
\bigg(\dfrac{\as(Q)}{\pi}\bigg)^3f_3^{\rm ang}(e)+\ldots\nonumber
\end{align}
Similarly we also expand the averaged and angular total cross-sections as
\begin{align}\label{eq:Rhad-expansion}
R_{\rm had} & = 1 + \dfrac{\as(Q)}{\pi}R_1 +
\bigg(\dfrac{\as(Q)}{\pi}\bigg)^2R_2+
\bigg(\dfrac{\as(Q)}{\pi}\bigg)^3R_3+\ldots\\
R_{\rm ang} & = \dfrac{\as(Q)}{\pi}R_1^{\rm ang}+
\bigg(\dfrac{\as(Q)}{\pi}\bigg)^2R_2^{\rm ang}+
\bigg(\dfrac{\as(Q)}{\pi}\bigg)^3R_3^{\rm ang}+\ldots\nonumber
\end{align}
Finally one can define the integrated or cumulant cross-section
\begin{align}\label{eq:cumulants}
\Sigma_{\rm ang}(e_c) & =
\dfrac{1}{\sigma_0}\int_{0}^{e_c}\df e\,\dfrac{\df \sigma_{\rm ang}}{\df e}\,,
\\
& = \dfrac{\as(Q)}{\pi}\,\Sigma_1^{\rm ang}(e_c) +
\bigg(\dfrac{\as(Q)}{\pi}\bigg)^2 \Sigma_2^{\rm ang}(e_c)+
\bigg(\dfrac{\as(Q)}{\pi}\bigg)^3 \Sigma_3^{\rm ang}(e_c)+\ldots\nonumber
\end{align}
At ${\mathcal O}(\as)$ we find the following analytic result for thrust
\begin{align}
\Sigma_1^{\rm ang}(\tau) = -\,\dfrac{3}{8}\,C_F
\Bigg[\dfrac{\tau\,(7-3\tau)}{1-\tau}+8\,\log(1-\tau)\Bigg]\,.
\end{align}
The expressions for arbitrary $\mu$ can be trivially obtained by expanding out $\as(Q)$
in terms of $\as(\mu)$ and $\log(\mu/Q)$.
The best strategy to obtain the two angular pieces is to directly project them out of the Event2 runs,
using the second and third lines of Eq.~(\ref{eq:final-general}). The corresponding integrals can be
performed event by event (that is, one does not need to make a two-dimensional grid in the event-shape
and $\cos\theta_T$ and integrate later). The cross-section of each event is simply weighted by either
$1 = P_0(\cos\theta_T^i)$ or $3/8\,(2-5\cos^2\theta^i_T) = 1/8 - 5/4\,P_2(\cos\theta_T^i)$
to project out the averaged and angular pieces respectively.
$\theta^i_T$ is the angle formed by the incoming electron and the thrust axis for the particular event
configuration. Since the projecting functions can be expressed in terms of linear combinations of
Legendre polynomials of order less than $3$, any hypothetical additional angular structure
expressed in terms of higher-order polynomial would be averaged out.
Finally events are clustered in histograms according to event-shape values.

As a cross check of the validity of Eq.~(\ref{eq:final-general}) we try to project out additional
angular structures. So we assume that there exist angular terms proportional to higher-order Legendre
polynomials. Since we only care about parity-conserving distributions only polynomials $P_n$ with
even $n$ can appear. We have projected out terms proportional to $P_4(\cos\theta_T)$ and
$P_6(\cos\theta_T)$, which are obtained by integrating with $9/2\,P_4(\cos\theta_T)$ and
$13/2\,P_6(\cos\theta_T)$, respectively. It is important to note that since 
$F_2(\cos\theta_T) = 1/2 + 1/8\,P_2(\cos\theta_T)$ and $F_1(\cos\theta_T) = -\,2\,P_2(\cos\theta_T)$
these two additional projections are not affected by the $F_0$ and $F_1$ terms. We found,
as expected, that the projected out terms are compatible with zero for all event-shapes.

\begin{figure*}[tbh!]
\subfigure[]
{
\includegraphics[width=0.475\textwidth]{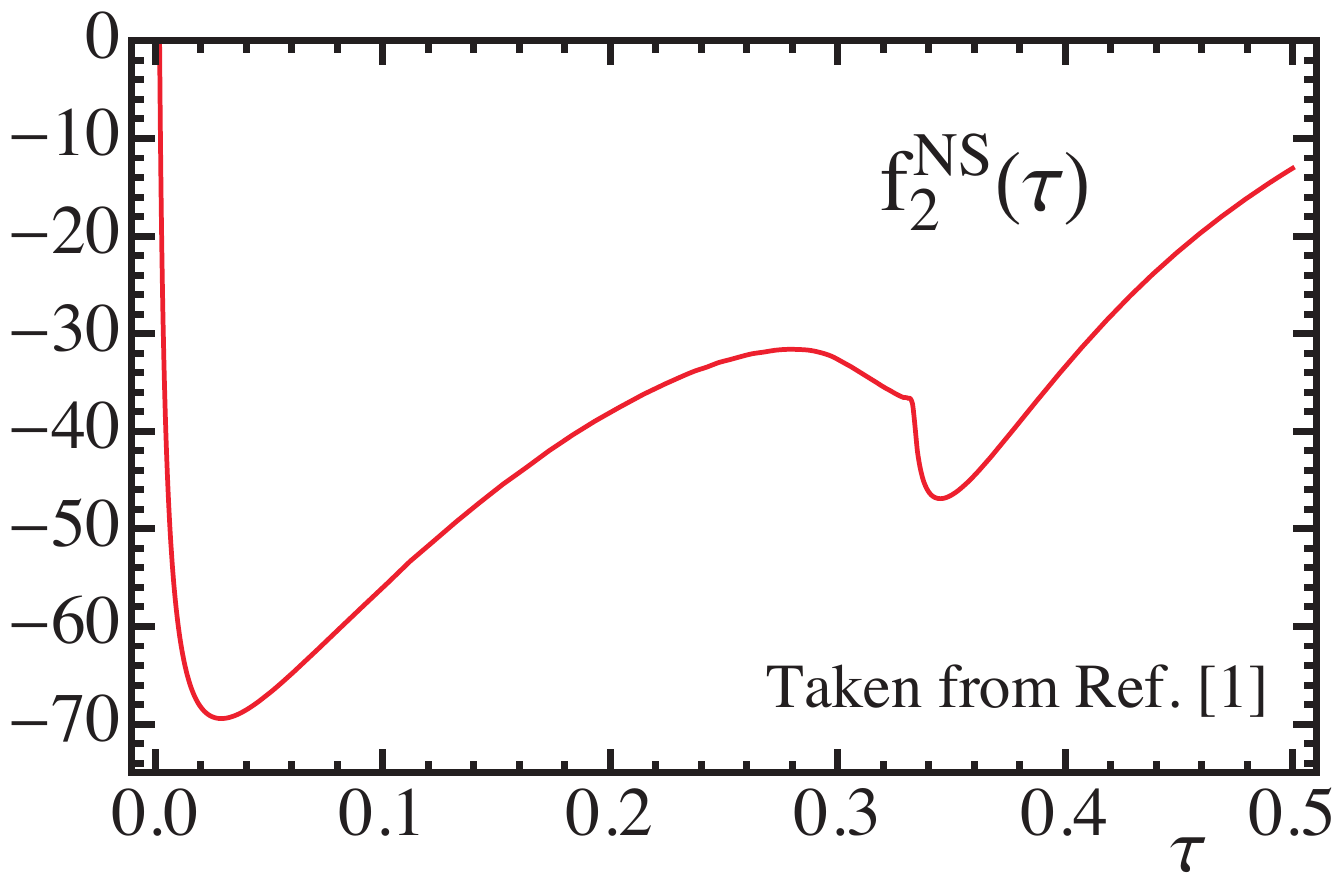}
\label{fig:NS-thrust}
}
\subfigure[]{
\includegraphics[width=0.485\textwidth]{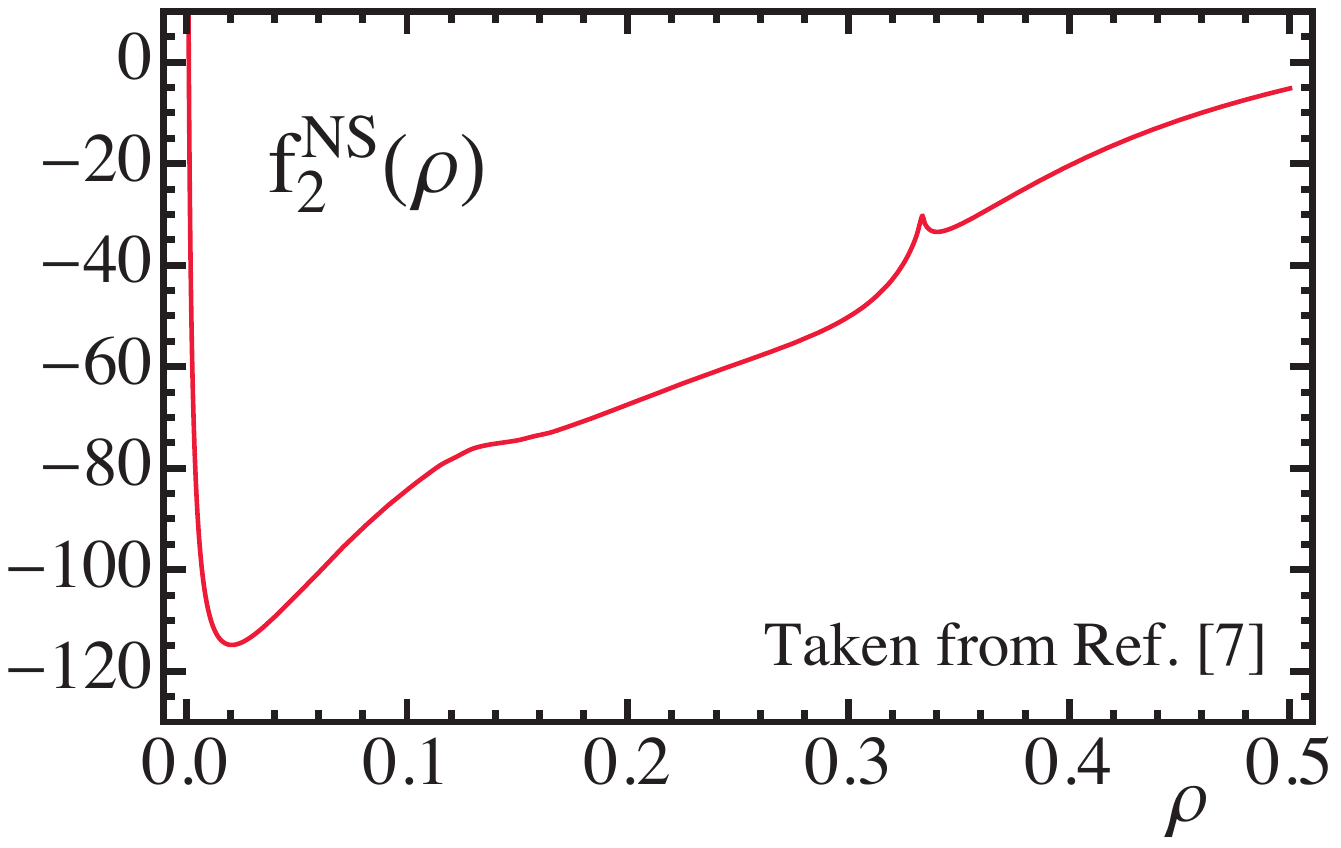}
\label{fig:NS-HJM}
}
\subfigure[]{
\includegraphics[width=0.475\textwidth]{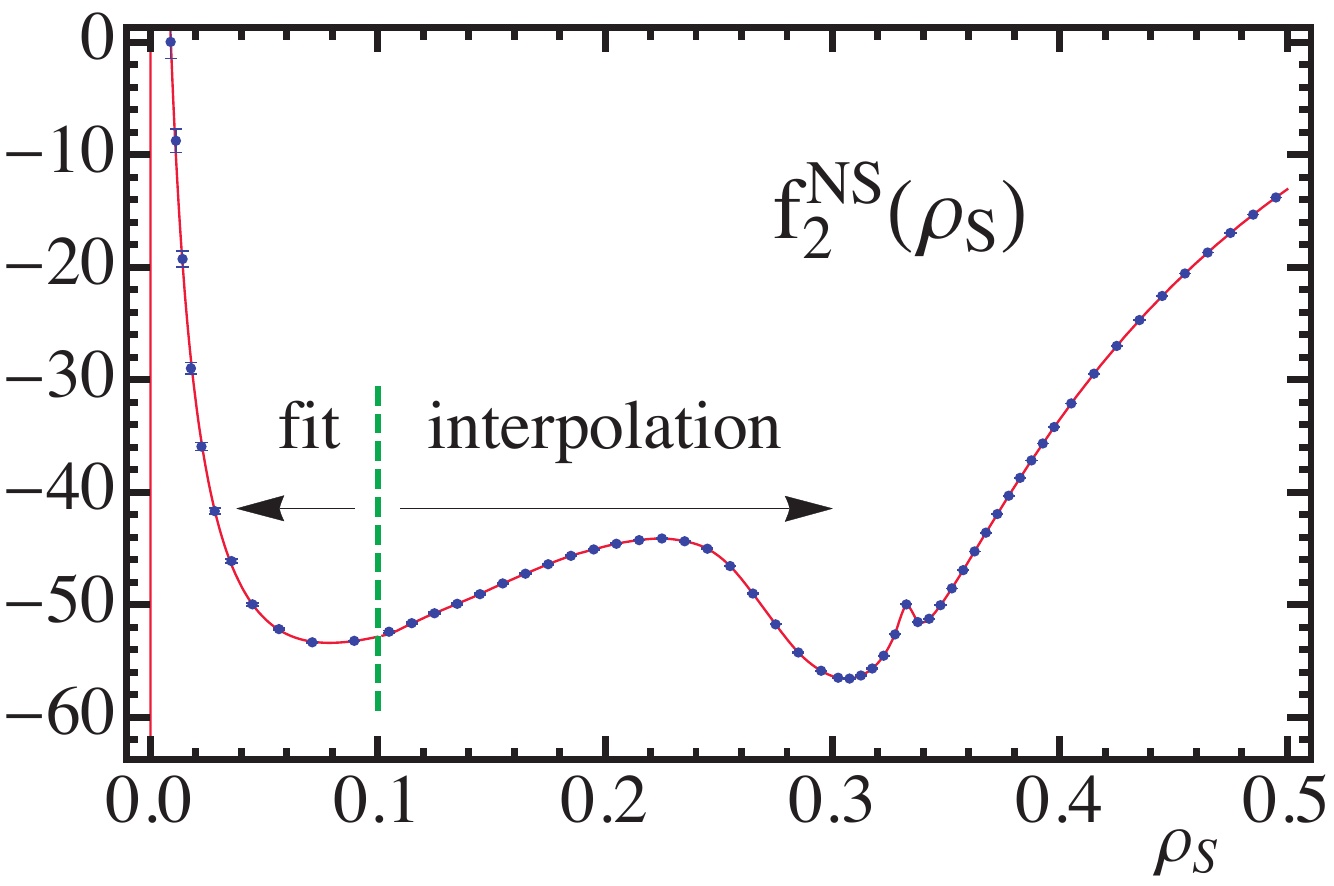}
\label{fig:NS-SJM}
}
\subfigure[]{
\includegraphics[width=0.475\textwidth]{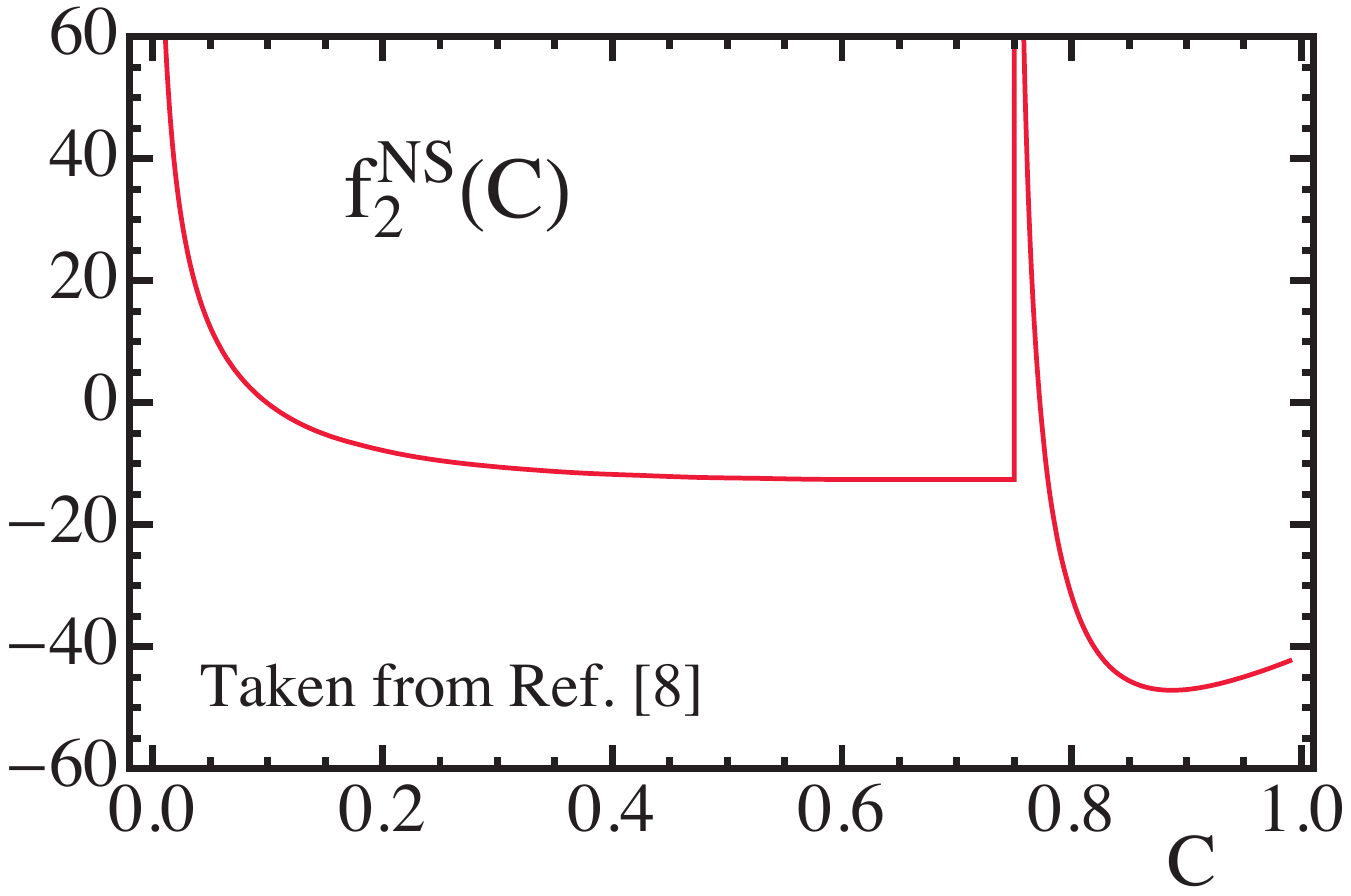}
\label{fig:NS-Cparam}
}
\caption{Non-singular distributions at $\mathcal{O}(\as^2)$ for (a) thrust, (b) Heavy-Jet Mass,
(c) the sum of Hemisphere Masses and (d) $C$-parameter. For the sum of Hemisphere Masses,
the blue dots correspond to Event2 output, linearly binned to the right of the green dashed line, and
logarithmically binned to the left. The red line corresponds to the function that we implement in our
numerical code. To the right of the dashed line we use an interpolating function, whereas to the left
we use a fit function. The error band corresponding to the fit function is too small to be visible in this
plot. For the other event shapes the non-singular function was determined in other publications.
\label{fig:SJM-NS}}
\end{figure*}

In Fig.~\ref{fig:NS-SJM} we show the extracted non-singular term for the sum of Hemisphere Masses. 
We directly compute the sum of all color structures, for the phenomenologically relevant case of five
light flavors. As it is well known, at the partonic level (that is for massless particles) thrust and 
the sum of Hemisphere Masses are identical in the dijet limit [\,see Eq.~(\ref{eq:MassesThrust})\,], 
and hence their singular distributions coincide to all orders in perturbation theory.
Non-singular terms are however different. The procedure to extract the non-singular terms is identical
to that followed in \cite{Abbate:2010xh}: we use a fit function below $\rho_S = 0.1$, which is fitted
to logarithmically binned Event2 data. Above $\rho_S = 0.1$, where errors are negligibly small, we use
an interpolation function over linearly binned Event2 data. For completeness we also show in
Fig.~\ref{fig:SJM-NS} the NLO
non-singular distributions for thrust, Heavy-Jet Mass and $C$-parameter.
\begin{figure*}[tbh!]
\subfigure[]
{
\includegraphics[width=0.475\textwidth]{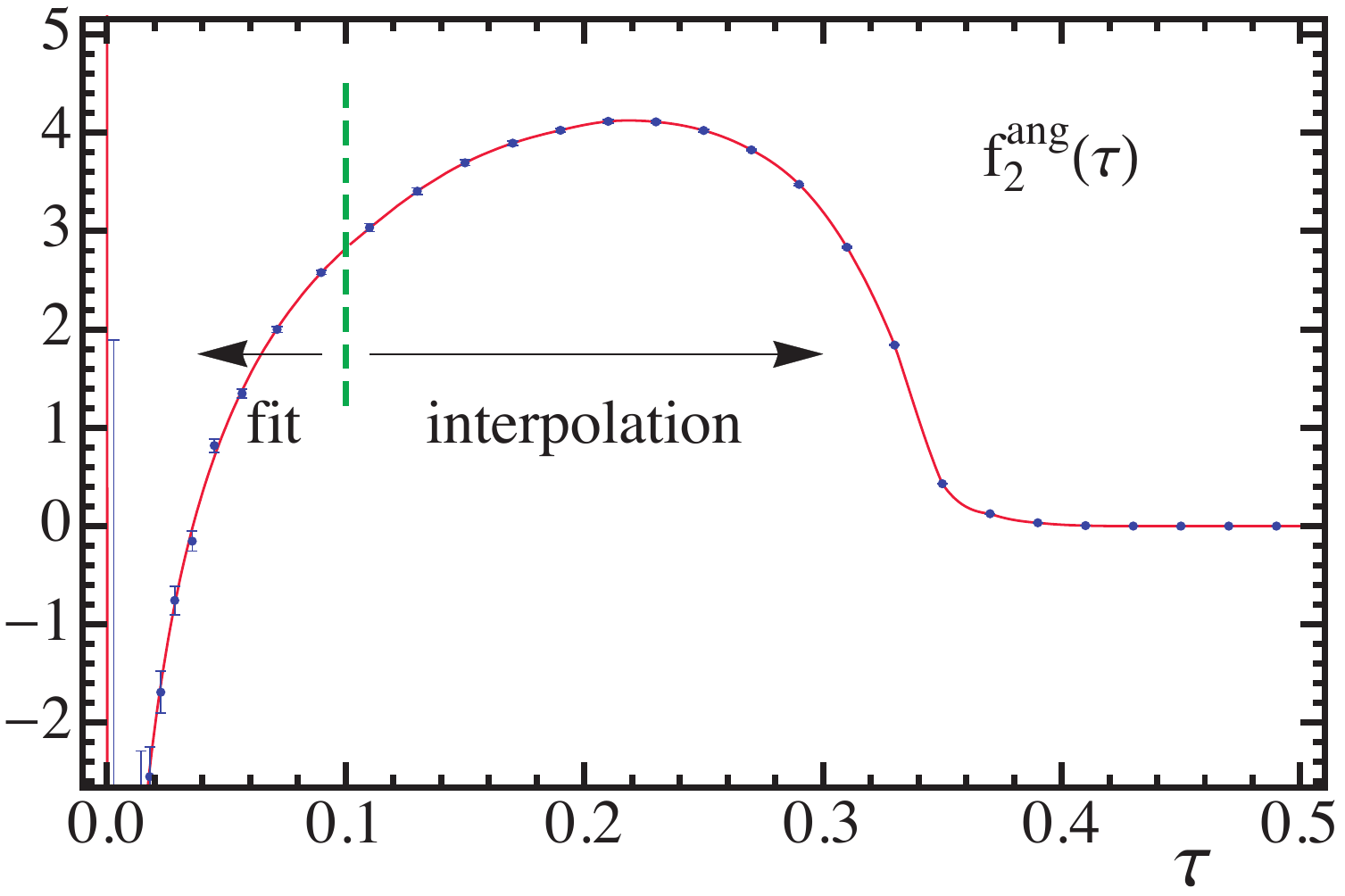}
\label{fig:angular-thrust}
}
\subfigure[]{
\includegraphics[width=0.475\textwidth]{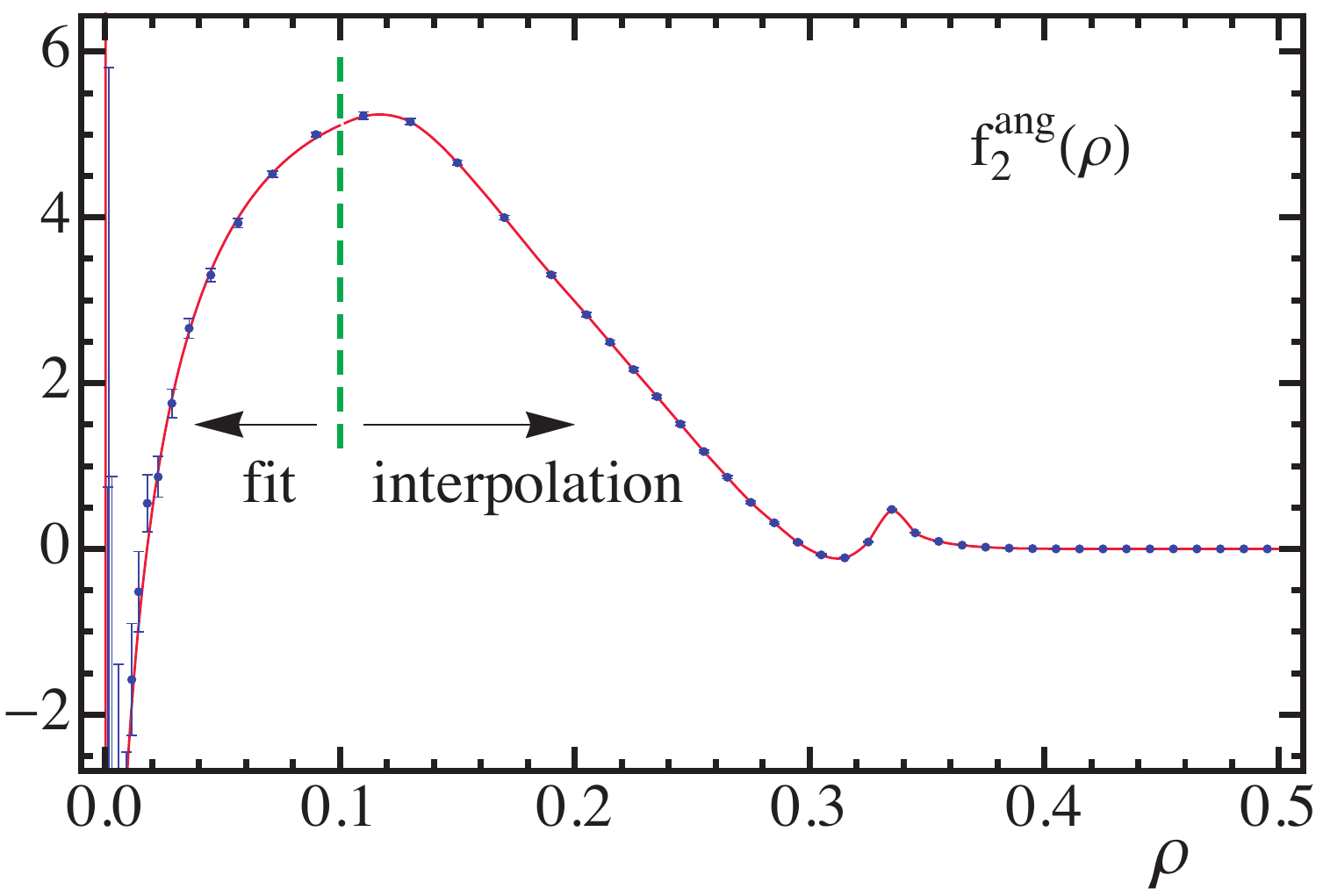}
\label{fig:angular-HJM}
}
\subfigure[]{
\includegraphics[width=0.475\textwidth]{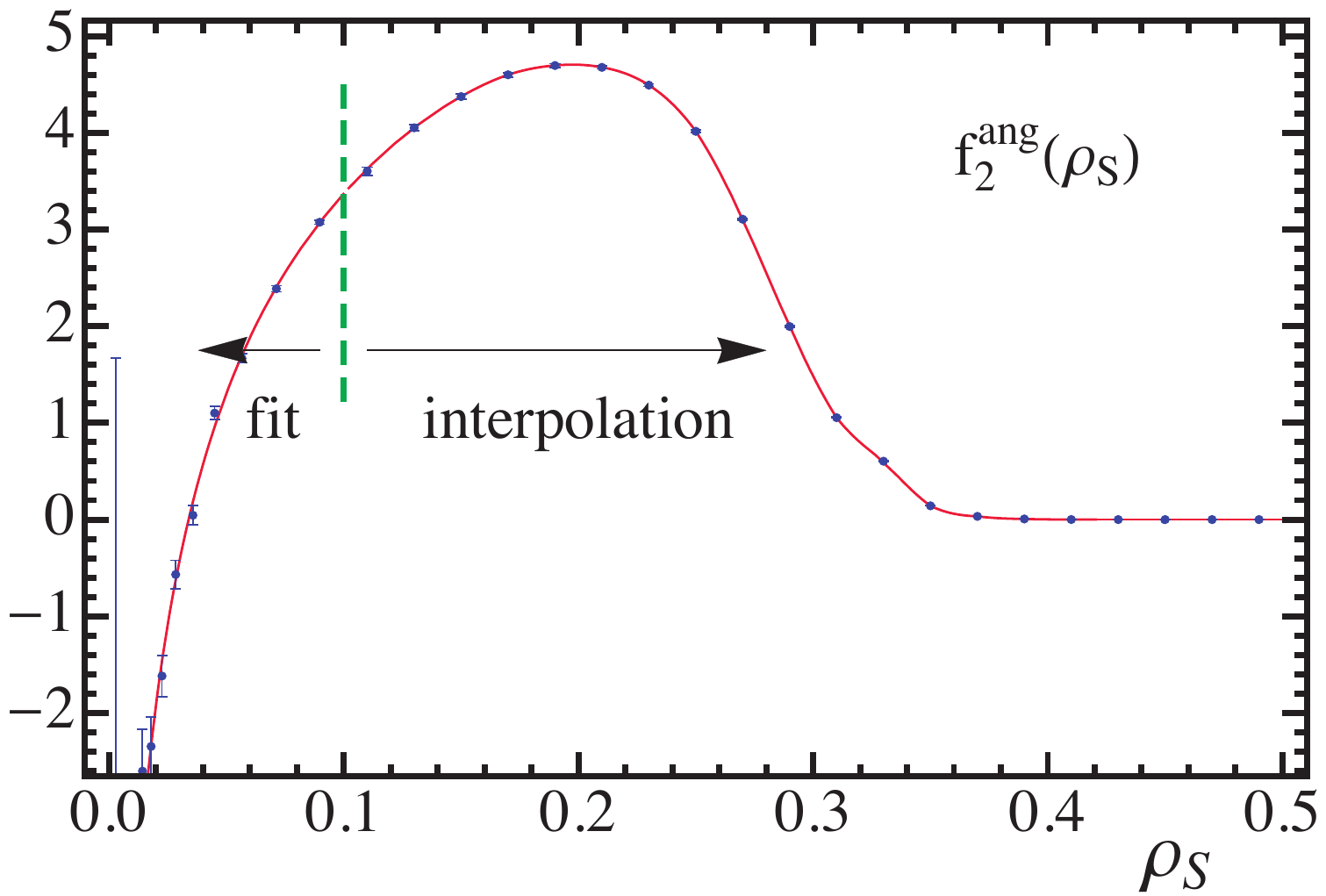}
\label{fig:angular-SJM}
}
\subfigure[]{
\includegraphics[width=0.475\textwidth]{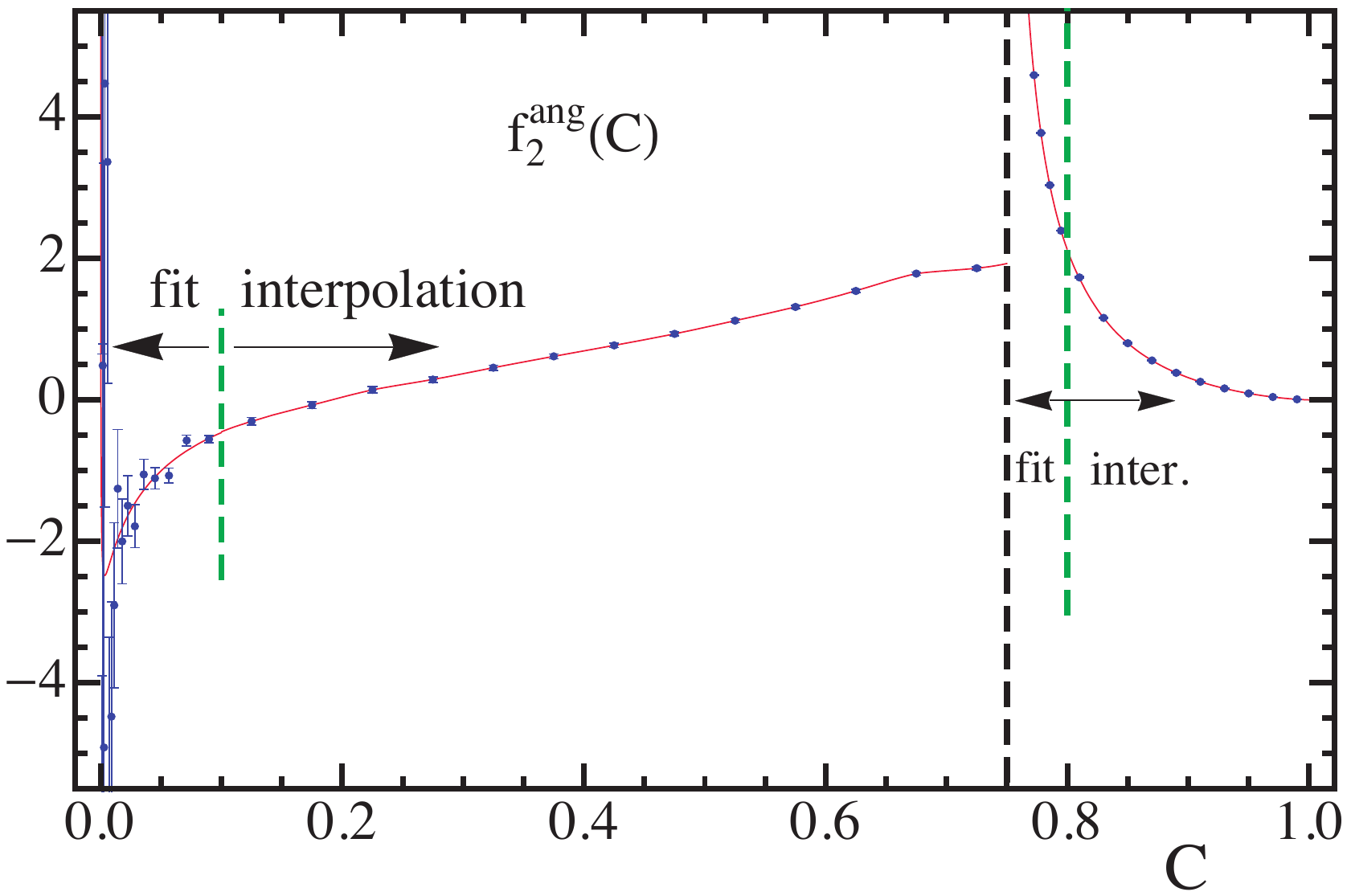}
\label{fig:angular-Cparam}
}
\caption{Angular distributions at $\mathcal{O}(\as^2)$. 
All the color structures with $n_f = 5$ are summed. The blue dots show the
Event2 output, linearly binned to the right of the green dashed line, and
logarithmically binned to the left. The red line corresponds to the determined function.
To the right of the dashed line we use an interpolating function, whereas to the right
we use a fit function. The error band corresponding to the fit function is too small to
be visible in this plot. The panels correspond to thrust (a); Heavy-Jet Mass (b), 
the sum of the Hemisphere Masses (c) and $C$-parameter (d).
In panel (d), the black dashed line shows the position of the ``shoulder'', which corresponds
to the four-particle threshold. We use a fit function and logarithmically binned Event2 output
between the green and black dashed lines, and an interpolation with linearly binned Event2
output above the second green line.
}
\label{fig:angular-distributions}%
\end{figure*}
\begin{figure*}[tbh!]
\subfigure[]
{
\includegraphics[width=0.475\textwidth]{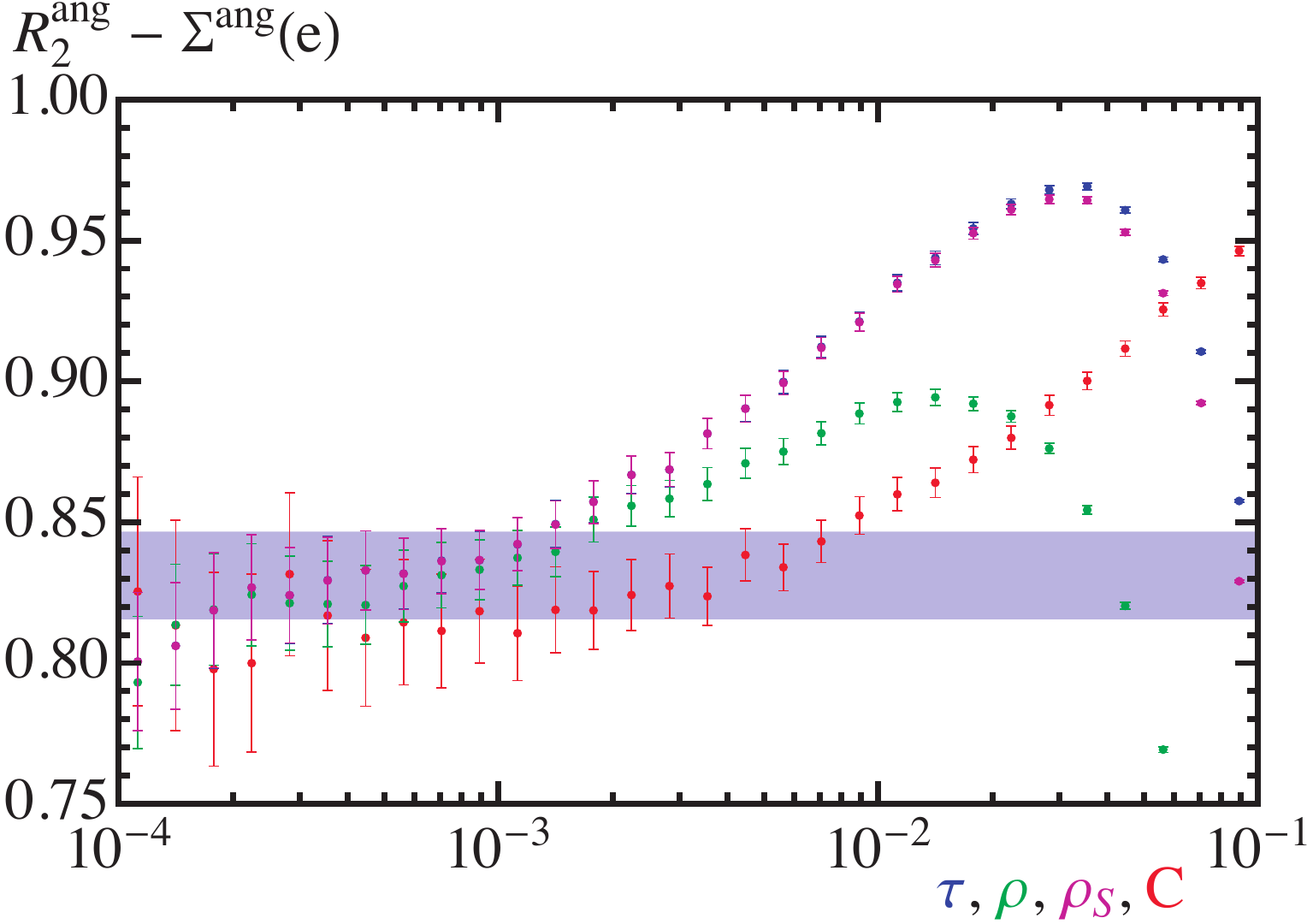}
\label{fig:total-All}
}
\subfigure[]{
\includegraphics[width=0.475\textwidth]{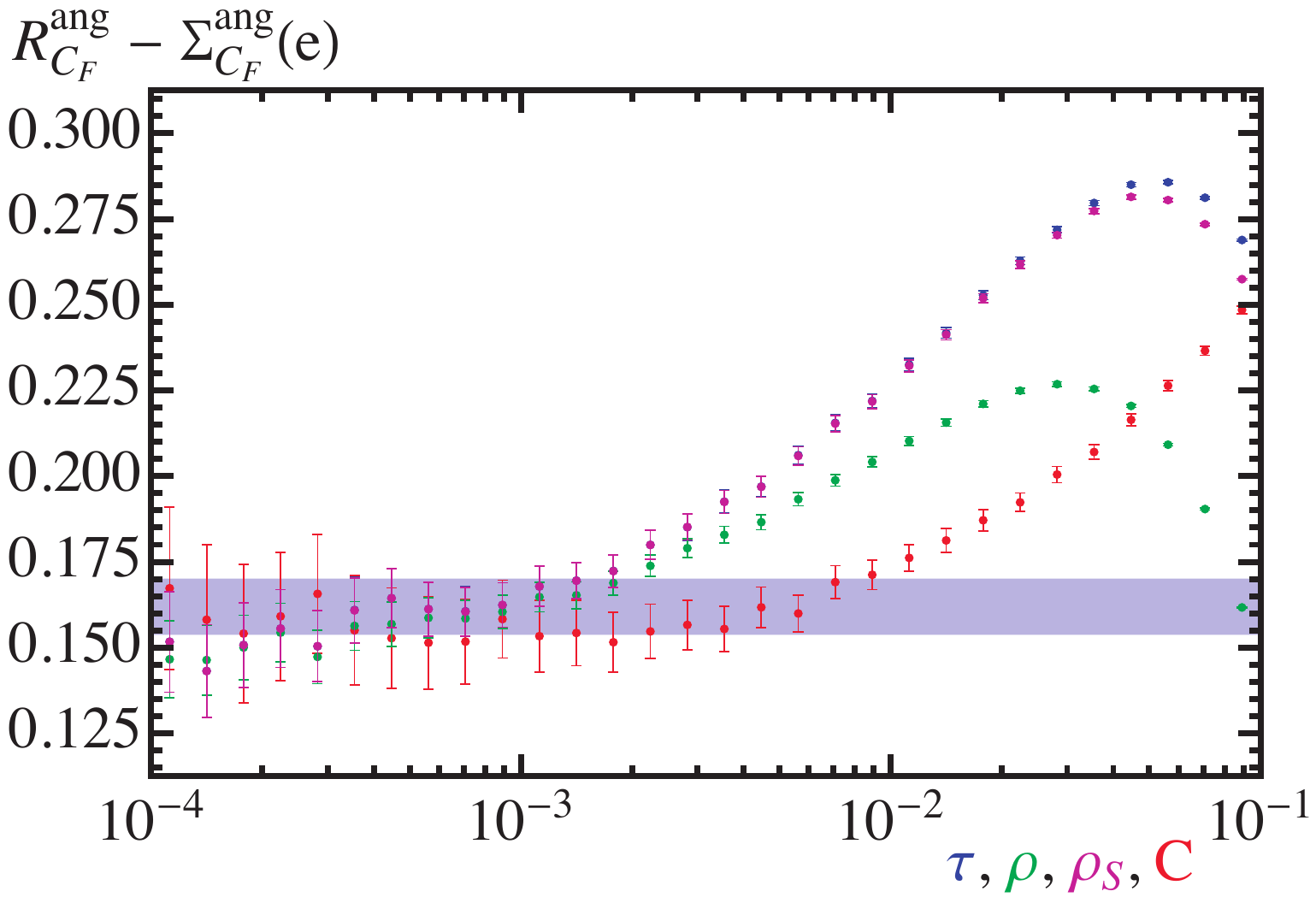}
\label{fig:total-CF}
}
\subfigure[]{
\includegraphics[width=0.475\textwidth]{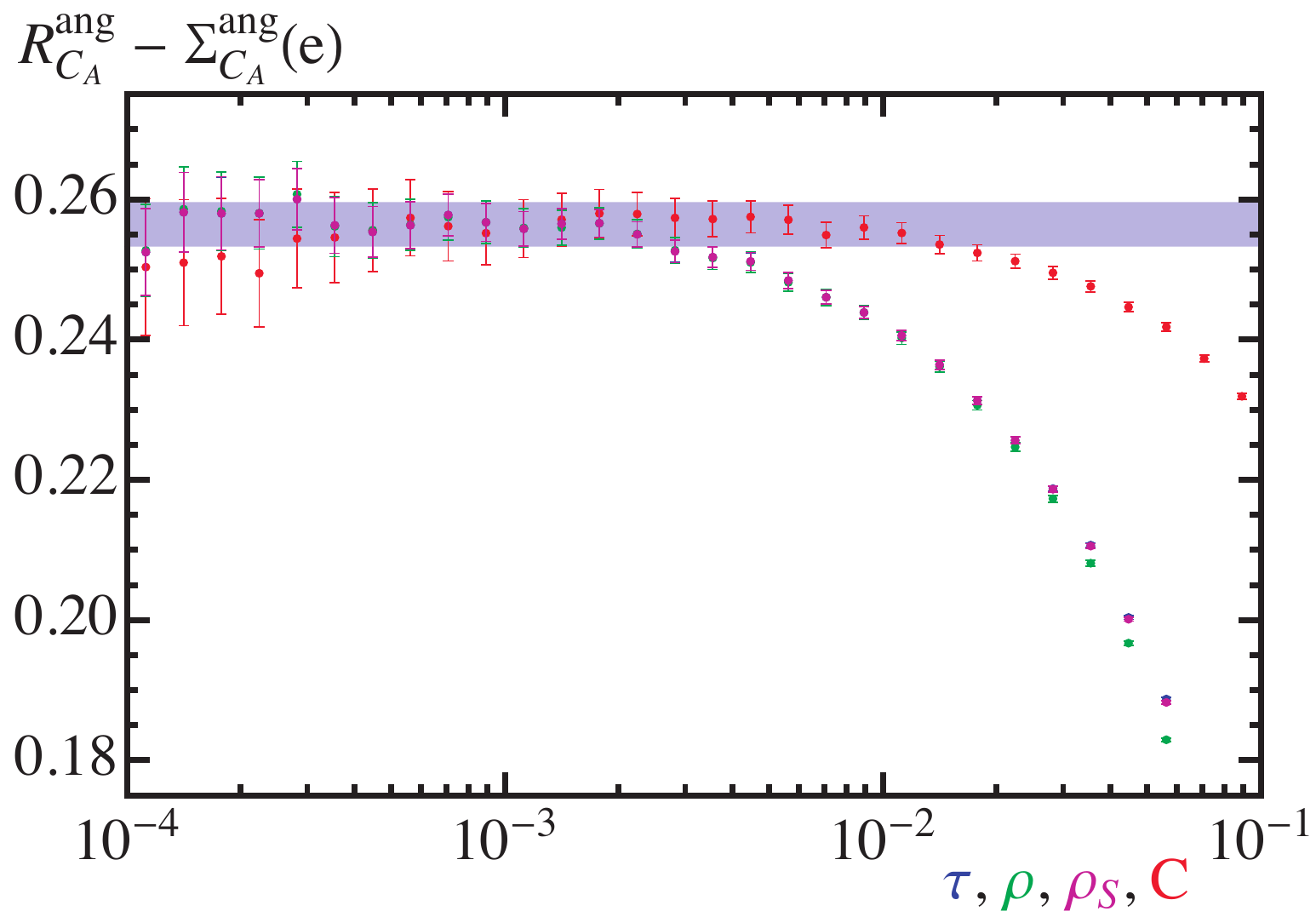}
\label{fig:total-CA}
}
\subfigure[]{
\includegraphics[width=0.475\textwidth]{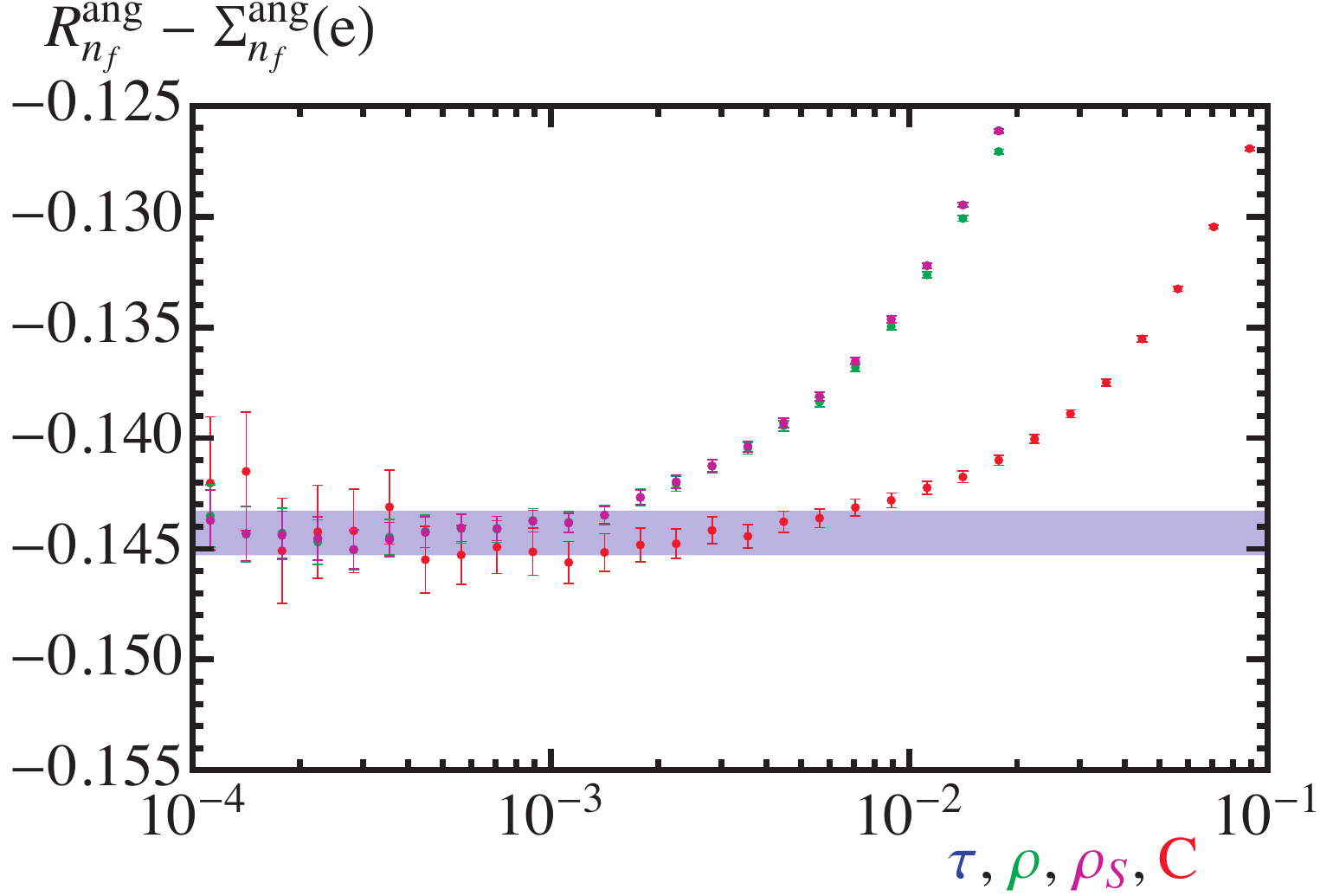}
\label{fig:total-Nf}
}
\caption{Computation of the angular piece for oriented total cross-section
[\,see Eqs.~(\ref{eq:angular-Rhad}), (\ref{eq:Rhad-expansion}) and 
(\ref{eq:Rang-split})\,] from Event2 output.
The cross-section is obtained by extrapolating to zero, which is simpler with 
logarithmic binning. The extrapolation is shown as a blue band. In each
plot we show various event-shape variables $e$: thrust in blue, $C$-parameter in red, Heavy-Jet Mass in green
and the sum of Hemisphere Masses in magenta. $\Sigma_2$ is the cumulant cross
section, and it is defined in
Eq.~(\ref{eq:cumulants}). In all cases, all event-shapes converge to the same value for
small values of the shape variable. In panel (a) the sum of the various color pieces for $n_f = 5$ is
shown; (b) shows the $C_F^2$ piece, (c) the $C_FC_A$ piece and (d) the $C_FT_f n_f$ term.}
\label{fig:total-X-section}
\end{figure*}
\begin{figure*}[tbh!]
\subfigure[]
{
\includegraphics[width=0.465\textwidth]{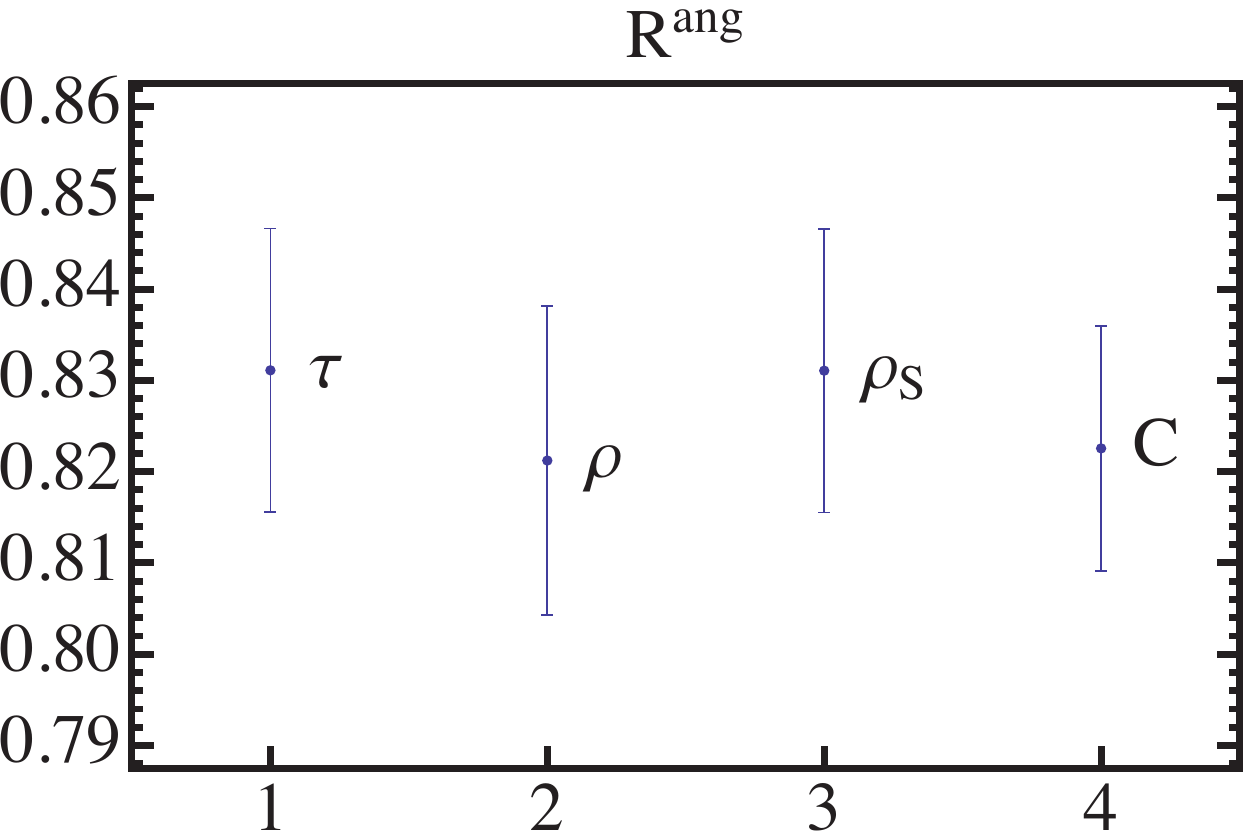}
\label{fig:X-total-All}
}
\subfigure[]{
\includegraphics[width=0.475\textwidth]{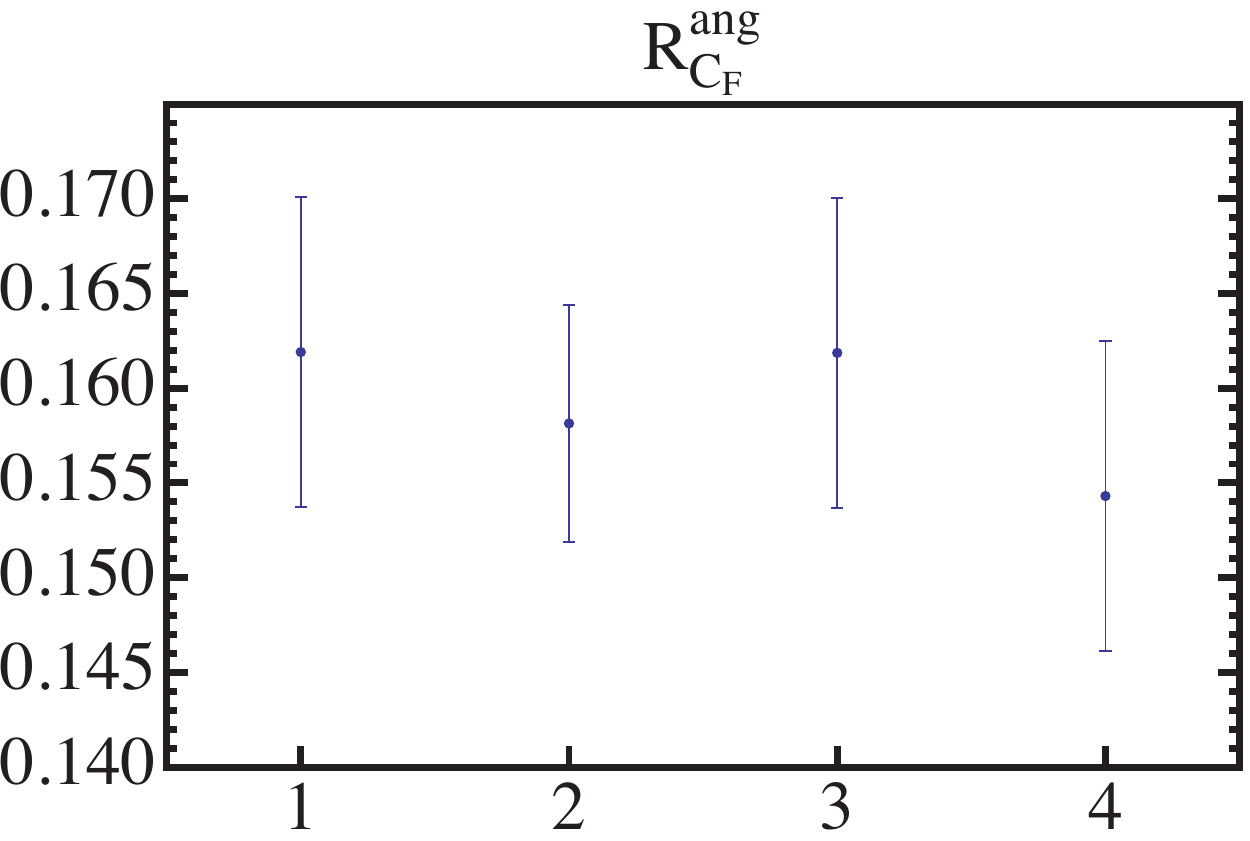}
\label{fig:X-total-CF}
}
\subfigure[]{
\includegraphics[width=0.465\textwidth]{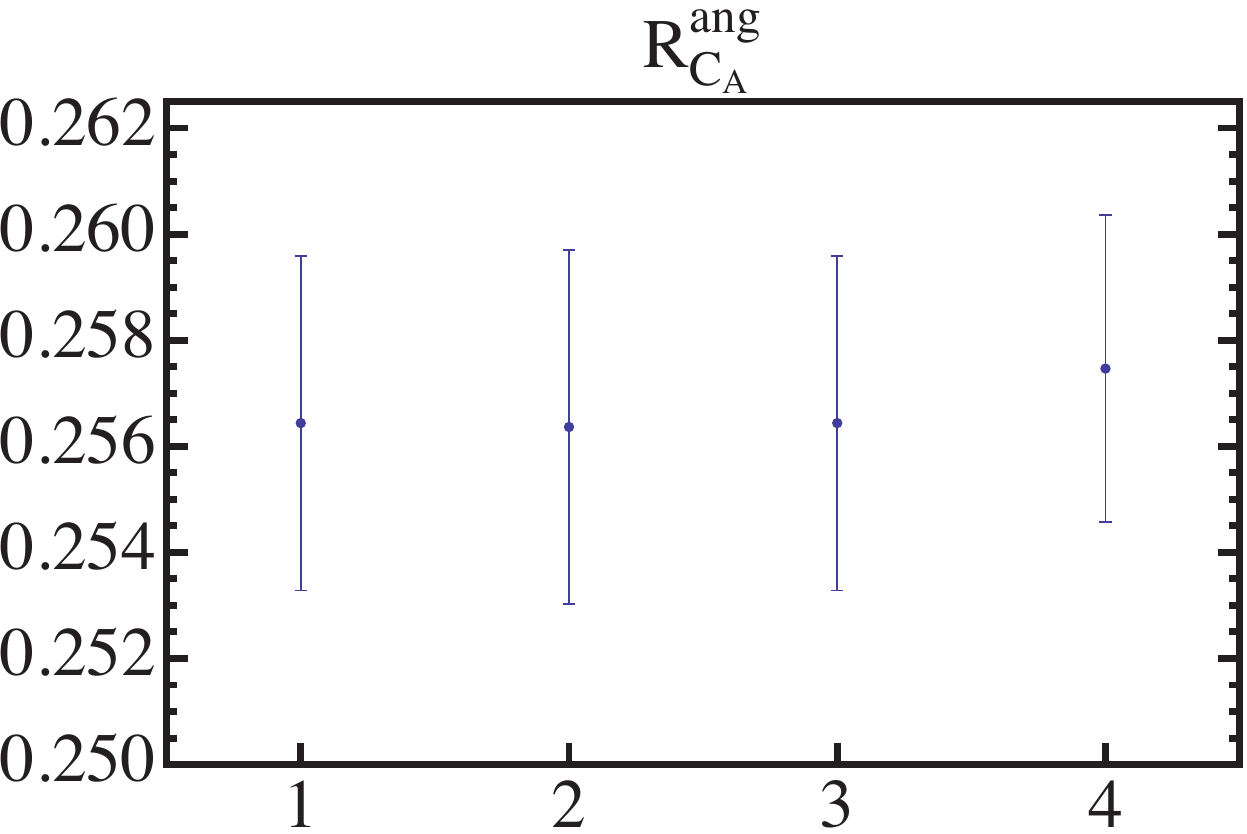}
\label{fig:x-total-CA}
}
\subfigure[]{
\includegraphics[width=0.475\textwidth]{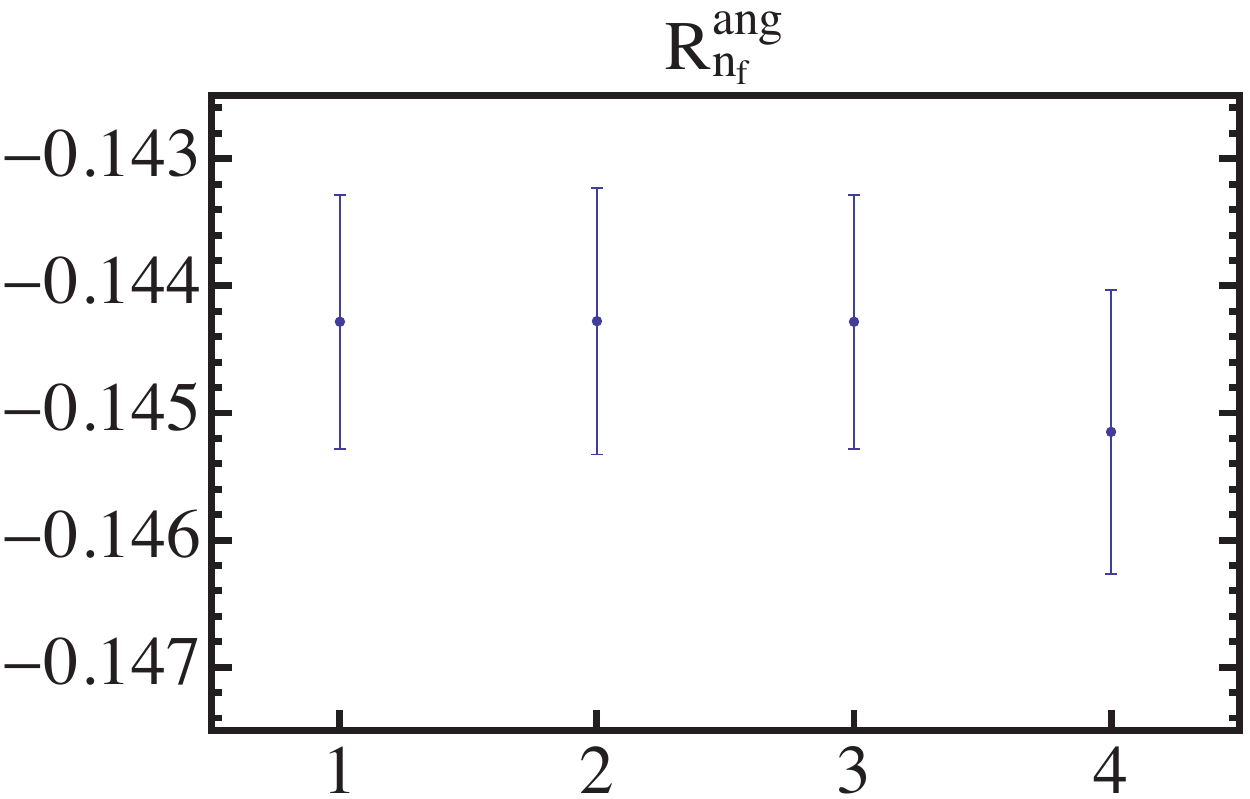}
\label{fig:X-total-Nf}
}
\caption{Comparison of the determination of the angular piece for oriented total cross-section.
On the horizontal axis $1$ corresponds to Thrust, $2$ to Heavy-Jet Mass, $3$ to the sum of the 
Hemisphere Masses, and $4$ to $C$-parameter. 
In panel (a) the sum of the various color pieces for $n_f = 5$ is
shown; (b) shows the $C_F^2$ piece, (c) the $C_FC_A$ piece and (d) the $C_FT_f n_f$ term.}
\label{fig:Comarison-total-X-section}%
\end{figure*}

In Fig.~\ref{fig:angular-distributions} we show the angular fixed-order distribution
at $\mathcal{O}(\as^2)$ for thrust (a), Heavy-Jet Mass (b), the sum of the
Hemisphere Masses (c) and $C$-parameter (d), as extracted from the Event2 program.
We have summed all color structures for $n_f = 5$.
Since the angular piece is numerically much smaller than the averaged one, the
relative errors are significantly larger. The strategy that we follow is the same
as for the non-singular terms of $\rho_S$, with a slight modification: the fit
function coefficients have an explicit dependence on the value of $R_2^{\rm ang}$,
which is known only numerically (see below). In this way, if we vary the value
of $R_2^{\rm ang}$ within errors, the fit function is varied accordingly, in such
a way that the total integral is always $R_2^{\rm ang}$, and the angular cross-section
is exactly normalized to one. The determination of the $C$-parameter deserves further
explanation. Since the ${\mathcal O}(\as)$ fixed-order distribution does not
fall off to zero in the completely symmetric configuration in which the three
partons have the same energy (that is for $C = 3/4$), the cross-sections has a
log-integrable singularity at ${\mathcal O}(\as^2)$, located precisely at
$C = 3/4$. This happens both for the averaged and the angular
distributions\,\footnote{Our analytical computation predicts
$f_1^{\rm ang}(3/4) = 8\,\sqrt{3}\,\pi\,C_F /81$.}\,.
This behavior is known as the shoulder, and a detailed explanation on
its physical origin and how to resum the corresponding logs can be found in
Ref.~\cite{Catani:1997xc}. We made a dedicated Event2 run for the region
above the shoulder, logarithmically binned around $C = 3/4$, that is we made
histograms in the variable $\log_{10}(C-3/4)$. We use a fit function for the
region $3/4 > C > 0.8$ using the log-binned output, and a fit function on
for $C > 0.8$ on linearly binned Event2 data.

The last piece of information that one needs to extract from Event2 is the
two-loop averaged total cross-section. Unfortunately one cannot simply sum
all of the randomly generated events, since Event2 discards events in which
partons are too close to each other, which means that the extreme dijet
region is not correctly described. This is clearly visible in the histograms
since for very small values of the event-shape errors are unnaturally large
and central values stop following a natural trend. What one has to do instead
is to sum up all events which produce values of a given event-shape $e$ bigger
that a small value $e_{\rm min}$, and then extrapolate to $e_{\rm min}\to 0$.
To do that we can use any event-shape. The simplest way is using a linearly
binned histogram, and linearly (or using a higher-degree polynomial)
extrapolate to zero using the last few points.
We discard this procedure because the extrapolation is affected by logarithms
[\,near zero the sum of bins behaves as $R_2^{\rm ang} + e\,\sum_i\log^i(e)$\,].
A better strategy is to use a logarithmically binned histogram. In this case
when approaching the dijet limit, the sum of histograms becomes exponentially
close to $R_2^{\rm ang}$. It is very simple to realize that this regimes has
been reached, since graphically the distribution becomes very flat (see
Fig.~\ref{fig:total-X-section}). This method was first applied
in Ref.~\cite{Hoang:2008fs}. To be more definite, we proceed as follows: first
we select a set of points that a) have central values which not show an increasing
or decreasing trend (they are flat within statistical fluctuations), this
cuts off points with too big value of the event-shape; and b)
are not yet affected by cutoff effects, which limits the points with very
small value of the event-shape. Given these points, we determine the central
value by averaging the central values (we do not make a weighted average
since errors are highly correlated). We determine a ``statistical'' uncertainty
by averaging the uncertainties of all the points. We assign a systematic
uncertainty by taking half of the maximum difference of central values.
We have used $\tau$, $\rho$, $\rho_S$ and $C$
to determine $R_2^{\rm ang}$ and we find very good agreement in each color
structure, as can be seen in Fig.~\ref{fig:Comarison-total-X-section}.

Defining
\begin{align}\label{eq:Rang-split}
R_2^{\rm ang} = C_F^2\, R^{\rm ang}_{C_{\!F}} + C_F C_A R^{\rm ang}_{C_{\!A}} +
C_F T_F\, n_f R^{\rm ang}_{n_{\!f}}\,,
\end{align}
we find
\begin{alignat}{7}
& R_2^{\rm ang}     &\;=&& 0.831     &&\;\pm&& 0.014  &&\;\pm&& 0.006\,,\\
& R^{\rm ang}_{C_{\!F}} &\;=&& 0.162     &&\;\pm&& 0.008  &&\;\pm&& 0.002\,, \nonumber\\
& R^{\rm ang}_{C_{\!A}} &\;=&& 0.256     &&\;\pm&& 0.029  &&\;\pm&& 0.001\,, \nonumber\\
& R^{\rm ang}_{n_{\!f}} &\;=&& \,-\,0.1443 &&\;\pm&& \;0.0008 &&\;\pm&& \;0.0006\,, \nonumber
\end{alignat}
where the first uncertainty is statistic and the second one is systematic.
Our results do not agree with those computed in Ref.~\cite{Lampe:1992au}. We find a much
bigger correction than they do, and additionally we find an opposite sign for $R^{\rm ang}_{C_{\!F}}$.
We will make additional checks of our determination in future work.
\begin{figure*}[t!]
\subfigure[]{
\includegraphics[width=0.475\textwidth]{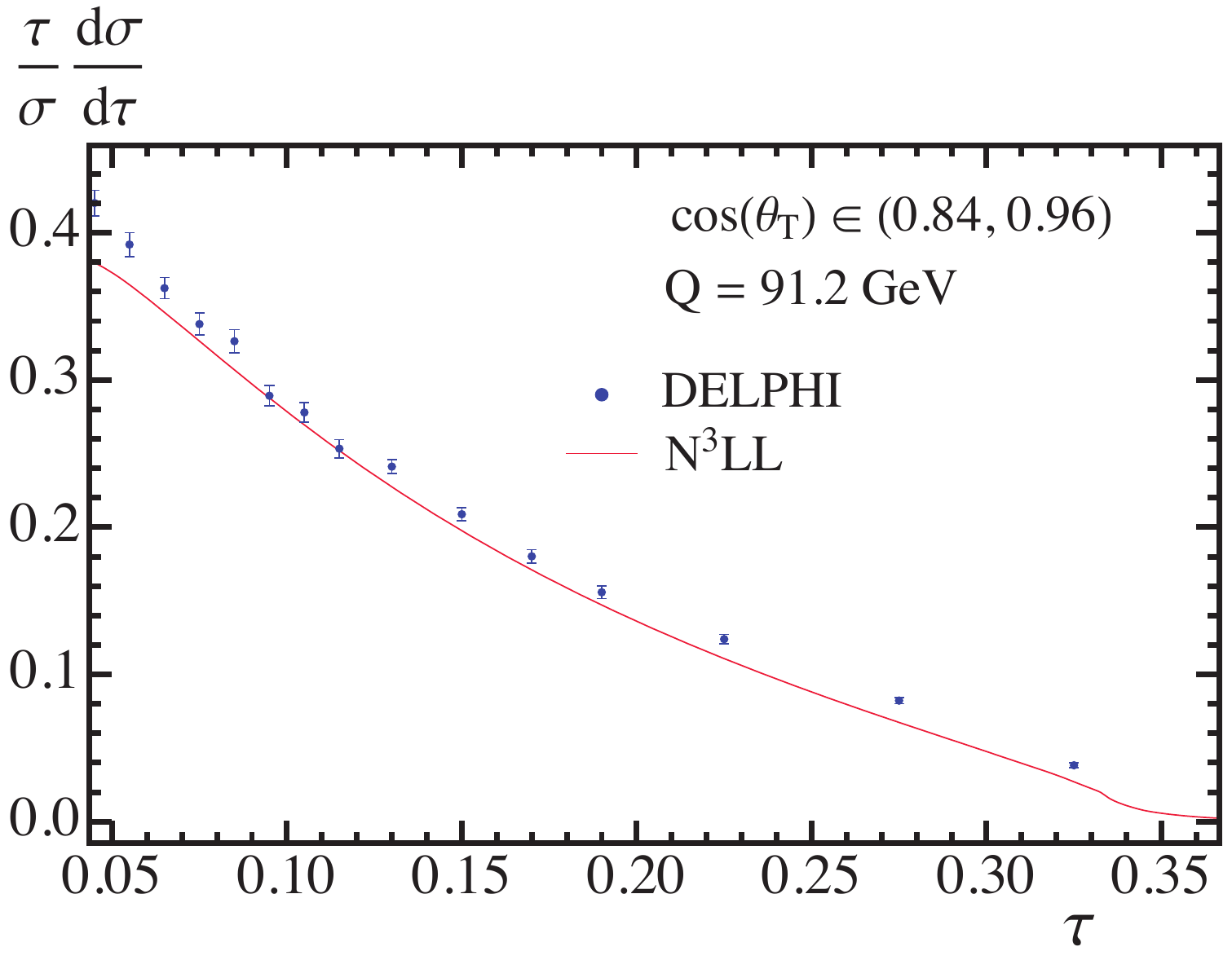}
\label{fig:Data-1}
}
\subfigure[]{
\includegraphics[width=0.475\textwidth]{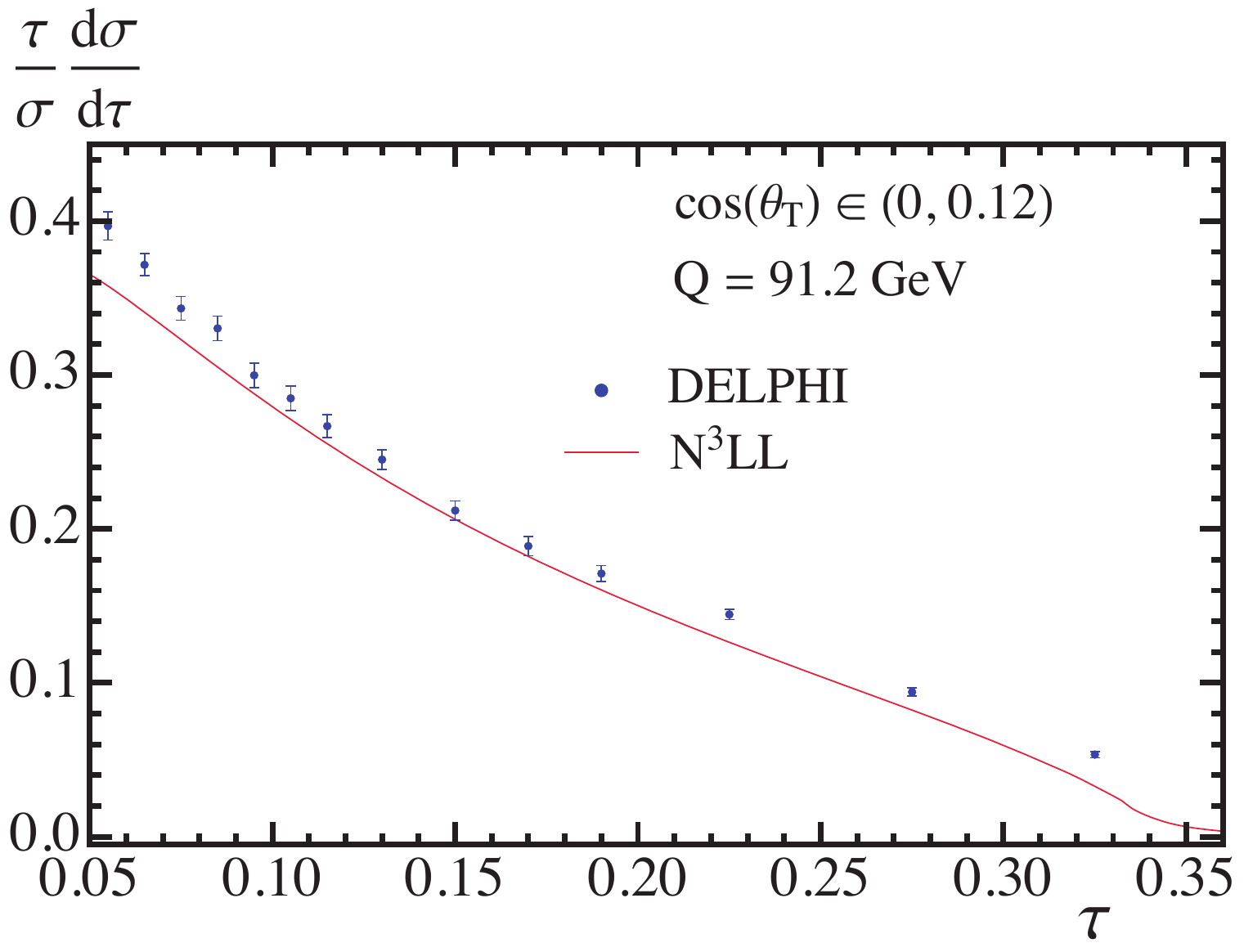}
\label{fig:Data-2}
}
\vspace{-0.2cm}
\caption{Comparison of our theoretical predictions (red line) with DELPHI data
(blue dots) for thrust. Our theoretical prediction contains resummation of singular logs at
N${}^3$LL and fixed-order matrix elements at $\mathcal{O}(\alpha_s^3)$ and
$\mathcal{O}(\alpha_s^2)$
for the total and angular distributions, respectively. No power corrections have been
included, which could explain the slight disagreement with data. We use the world average
value for $\alpha_s(m_Z)$. We compare to the two bins in $\cos \theta_T$ for which the
difference between the averaged and oriented distributions is maximal. In panel (a) the
oriented distribution is higher than the averaged, whereas in panel (b) the opposite occurs.
\label{fig:Data}}
\end{figure*}

In Fig.~\ref{fig:Data} we compare our theoretical predictions with DELPHI data. We compare
the differential thrust distribution for two bins in $\cos \theta_T$. We choose these two
bins since they have the largest deviation from the averaged cross section in the positive
and negative directions. Our theoretical prediction is purely perturbative, and includes
resummation of singular logs at N${}^3$LL and fixed-order matrix elements up to three loops.

\section{Conclusions}
\label{sec:conclusions}
We have performed a first theoretical analysis of oriented event-shape observables,
which are double differential distributions in the event-shape 
variable and in the polar angle of the thrust axis with respect to 
the electron--positron beam, in the context of SCET.
The event-shape variables analyzed were thrust, 
Heavy-Jet Mass, the sum of the Hemisphere Masses and \mbox{$C$-parameter}.
We have proven that perturbation theory predicts that the angular 
dependence of the oriented event-shapes is parameterized in terms 
of only two angular structures, $F_0 = 3/8\,(1 + \cos^2\theta_T)$ and 
$F_1 = (1 - 3\,\cos^2\theta_T)$. The most singular contributions, 
as predicted by SCET, inherit the angular dependence of the lowest 
order process $e^+e^-\to q\bar q$, and therefore can arise exclusively 
in the term which is proportional to $F_0$. This general behavior 
is extensible to recoil-sensitive variables such as Jet Broadening 
because only the hard function is sensitive to the orientation of 
the thrust axis, and also from the partonic to the hadron level.  

We have extracted from fixed-order calculations the non-singular 
contributions to the angular averaged cross-section, which is 
proportional to $F_0$, and the new angular contribution of the 
$F_1$ term, at $\mathcal{O}(\as)$ analytically and at
$\mathcal{O}(\as^2)$ with the program Event2.
These are all the ingredients necessary to perform the determination of 
the strong coupling $\as$ from oriented event-shapes with resummed 
theoretical predictions at N${}^3$LL and ${\cal O}(\as^2)$ accuracy. 

The validity of the proof in Sec.~\ref{sec:Angular} can be extended to other axes other than the
thrust one. In principle any axis which is defined as the sum of the 3-momenta of particles is
equally valid, as this would allow to factor the phase space in a way analogous to
Eq.~(\ref{eq:phase-space-factor}). For instance, the momentum of the hardest particle, or the
momentum of the hardest jet, as long as the momenta of the jet is the sum of the momenta of
some of the particles it contains, are valid axes.

If LEP data are preserved \cite{Holzner:2009ew,Akopov:2012bm} at the particle level it could be
reanalyzed to produce very accurate oriented-event shape distributions at energies other than
the $Z$-pole. Actually it would be possible to directly determine the averaged and angular distribution
using the projection procedure sketched in Eq.~(\ref{eq:final-general}).
These results would be useful also for the measurement of $\as$ 
at higher energies at a future Linear Collider. 

\section*{Acknowledgments}
This work has been supported by the Research Executive Agency (REA) of the European Union under 
the Grant Agreement number PITN-GA-2010-264564 (LHCPhenoNet),
by the Spanish Government and EU ERDF funds (grants FPA2007-60323, FPA2011-23778 and CSD2007-00042 
Consolider Project CPAN) and by GV (PROMETEUII/2013/007).
VM acknowledges support from Marie Curie actions (PIOF-GA-2009-251174).
We thank the Erwin Schr\"odinger Institute (ESI) in Vienna for the warm
hospitality while part of this work was completed.

\bibliography{thrust3}
\bibliographystyle{JHEP}

\end{document}